\documentclass[useAMS,usenatbib]{mn2e}   

\usepackage[dvips]{graphicx}
\usepackage{amssymb}
\usepackage{amsmath}
\usepackage{amstext}

\newcommand{\HI}{H\,{\sc i}}
\newcommand{\HII}{H\,{\sc ii}}

\newcommand{\Ha}{H$\alpha$}

\newcommand{\kms}{~km\,s$^{-1}$}

\newcommand{\FHI}{$F_{\rm HI}$}

\newcommand{\LB}{$L_{\rm B}$}
\newcommand{\LBo}{$L_{\odot}$}

\newcommand{\MHI}{$M_{\rm HI}$}

\newcommand{\Mo}{~M$_{\odot}$}
\newcommand{\Msun}{~M$_{\odot}$}
\newcommand{\Lo}{~L$_{\odot}$}
\newcommand{\Lsun}{~L$_{\odot}$}
\newcommand{\Zo}{~Z$_{\odot}$}

\newcommand{\AB}{A$_{\rm B}$}

\newcommand{\abox}{12+log(O/H)}
\newcommand{\Myr}{Myr}

\newcommand{\nodata}{...}
\newcommand{\Mdyn}{$M_{\rm dyn}$}

\newcommand{\MHil}{$M_{\rm H\, I}/L_B$}

\newcommand{\Mstars}{$M_{\rm stars}$}
\newcommand{\Mg}{$M_{\rm gas}$}

\newcommand{\sfrha}{$SFR_{\rm H\alpha}$}

\newcommand{\sfrfir}{$SFR_{\rm FIR}$}

\newcommand{\sfrghz}{$SFR_{\rm 20cm}$}
\newcommand{\sfrfuv}{$SFR_{\rm FUV}$}
\newcommand{\Moy}{~M$_{\odot}$\,yr$^{-1}$}
\newcommand{\MMoy}{M$_{\odot}$\,yr$^{-1}$}

\title[Intriguing \HI\ gas in NGC 5253]
      {The intriguing \HI\ gas in  NGC 5253: \\ 
      an infall of a diffuse, low-metallicity \HI\ cloud?\thanks{
       The observations were obtained with the Australia Telescope which is 
       funded by the Commonwealth of Australia for operations as a National 
       Facility managed by CSIRO.}}

\author[\'A.R. L\'opez-S\'anchez et al.]
       {\'A.R. L\'opez-S\'anchez$^{1,2,3}$\thanks{E-mail: alopez@aao.gov.au}, B.S. Koribalski$^3$,  
       J. van Eymeren$^4$, C. Esteban$^{5,6}$,
	\newauthor 
	 E. Kirby$^{7,3}$, H. Jerjen$^7$  \& N. Lonsdale$^8$ \\ 
	       $^1$Australian Astronomical Observatory, PO BOX 296, Epping, NSW 1710, Australia\\
               $^2$Department of Physics and Astronomy, Macquarie University, NSW 2109, Australia\\
               $^3$CSIRO Astronomy and Space Science / Australia Telescope National Facility, PO BOX 76, Epping, NSW 1710, Australia\\
        $^4$Universitaet Duisburg-Essen, Faculty of Physics, Lotharstr. 1, 47048 Duisburg, Germany\\
	$^5$Instituto de Astrof\'\i sica de Canarias, E-38200 La Laguna, Tenerife, Spain \\
	$^6$Departamento de Astrof\'{\i}sica, Universidad de La Laguna, E-38205 La Laguna, Tenerife, Spain\\
	$^7$Research School of Astronomy and Astrophysics, Australian National University, Cotter Rd. Weston, ACT 2611, Australia\\
	$^8$Centre for Astrophysics and Supercomputing, Swinburne University of Technology, PO Box 218, Hawthorn, Victoria 3122, Australia
}

\date{Received date: 18 Mar 2011; accepted date: 1 Sep 2011}

\begin{document}

\maketitle


\begin{abstract}
We present new, deep \HI\ line and 20-cm radio continuum data of the very puzzling blue compact dwarf galaxy 
NGC~5253, obtained with the Australia Telescope Compact Array 
as part of the `Local Volume \HI\ Survey' (LVHIS).
Our low-resolution \HI\ maps show, for the first time, the disturbed \HI\ morphology that NGC~5253 possesses, including tails, plumes and detached \HI\ clouds. 
The high-resolution map reveals an \HI\ plume at the SE 
and an \HI\ structure at the NW that surrounds an \Ha\ shell.
This latter structure is related to an expanding bubble in the ISM, 
but it will almost certainly not originate a galactic wind. 
We confirm that the kinematics of the neutral gas in NGC~5253 are highly perturbed and do not follow a rotation pattern. 
We discuss the outflow and infall scenarios to explain such disturbed kinematics, 
analyze the environment in which NGC~5253 resides, and compare it properties 
with those observed
in similar star-forming dwarf galaxies.  
The radio-continuum emission of NGC~5253 is resolved and associated with the intense 
star-forming region located at the center of the galaxy. 
We complete the analysis using multiwavelength data extracted from the literature, 
which includes \mbox{X-ray,} GALEX far-ultraviolet, 
optical $B$- and $R$-band and \Ha, near-infrared 
 \mbox{$H$-band,} and far-infrared data. 
We estimate the star-formation rate 
using this multiwavelength approach, 
and compare the results with other galaxy properties. 
NGC~5253 does not satisfy the Schmidt-Kennicutt law of star-formation, has a very low \HI\ mass-to-light ratio when comparing with its stellar mass, and seems to be slightly metal-deficient in comparison with starbursts of similar baryonic mass.
Taking into account all available multiwavelength data, we conclude that NGC~5253 is probably experiencing the infall of a diffuse, low-metallicity \HI\ cloud along the minor axis of the galaxy, which is comprising the ISM and triggering the powerful starburst. The tidally disturbed material observed at the east and north of the galaxy is a consequence of this interaction, which probably started more than 100~\Myr\ ago. The origin of this \HI\ cloud 
may be related with a strong interaction between NGC~5253 and the late-type spiral galaxy M~83 in the past.
\end{abstract}


\section{Introduction} 

Multiwavelength analyses of local dwarf galaxies give the opportunity to study 
the impact of the star-formation activity on both the interstellar medium (ISM) 
and the surrounding intergalactic medium (IGM),
the dynamics of galaxy evolution at both high and low spatial scales,
and explore the triggering mechanisms of low-mass starbursting systems.
In particular, nearby dwarf starbursts are excellent laboratories to investigate the properties 
of the star-forming regions, the behavior and fate of the neutral, ionized and hot gas, 
and their effects on the stellar component of the galaxy, giving fundamental parameters 
to constrain models of the chemical and dynamical evolution of galaxies.

An intriguing local starburst galaxy
is the blue compact dwarf galaxy (BCDG) 
NGC~5253 in the Centaurus A group \citep{Karachentsev07}. The larger nearby galaxy to NGC~5253 is the late-type spiral M~83, which is at a distance of  
4.6~Mpc and lies at an angular distance of $1^{\circ}$\,54' from NGC~5253.
NGC~5253 is an ideal target for multiwavelength analyses of starburst galaxies, and a large amount of information is available at all frequencies.
The outer optical isophotes of NGC 5253 resemble a dwarf elliptical galaxy, 
but \Ha\ images \citep{Martin1998,Calzetti04,Meurer06} 
soon revealed its starbursting nature,
as it shows a large amount of ionized gas. This includes a filamentary structure that is extending perpendicular 
to its optical major axis and reaches beyond the stellar distribution. 
The star-formation activity is very intense in this galaxy. Indeed,
NGC~5253 is considered as one of the youngest starbursts of the Universe 
\citep*[e.g.,][]{vdBergh80,MoorwoodGlass82, RiekeLW88,CaldellPhillips89}
and even a large population of massive, young Wolf-Rayet (WR) stars has been reported in this galaxy 
\citep*{CTM86, Schaerer97, Kobulnicky97, LSEGRPR07,LSE10a,LSE10b,Monreal-Ibero+10}.
The winds of the WR stars seem to be polluting some internal areas of the galaxy center, as a localized nitrogen enrichment 
\citep{WR89,Kobulnicky97, LSEGRPR07,LSE10a,LSE10b,Monreal-Ibero+10} 
and a possible slight helium enrichment \citep{CTM86, LSEGRPR07}
have been reported.
The highest oxygen abundance of the ionized gas found in NGC~5253 is 12+log(O/H)=8.30 \citep{LSEGRPR07}, 
that corresponds to $\sim$40\% solar assuming 12+log(O/H)$_{\odot}$=8.66 \citep*{ASP05}. 
The very center of the galaxy hosts an extremely luminous, compact, obscured \citep{Alonso-Herrero04} 
site of star-formation, including some star clusters with ages younger than 10~Myr and masses between 10$^5$ and 10$^6$~\Mo\ 
\citep*[e.g.,][]{Gonzalez-RiestraRZ87, Calzetti97, Vanzi04, Harris04, Martin-Hernandez05,Cresci05}.
Extended diffuse X-ray emission \citep{StricklandStevens99,Summers04} 
associated with the filamentary \Ha\ structure are explained as superbubbles around the massive star-forming regions, and have been created by the combined action of supernova explosions and stellar winds. However, the radio spectral energy distribution of NGC~5253 is almost entirely dominated by thermal emission 
\citep*{Beck96, THB98}, 
that implies the extreme youth of the starburst.


\begin{table*} 
\caption{Basic properties of the BCDG NGC~5253 and the dwarf irregular galaxies ESO~444-G084, IC~4316 and NGC~5264.}
\label{basic} 
\begin{tabular}{l ccc cc}
\hline
Name                    & NGC~5253    & IC~4316   &  ESO~444--G084     &    NGC~5264   & Ref. \\
HIPASS                  & J1339--31A  & J1340--28 &  J1337--28        &    J1341--29  &   \\

\hline
center position        & $13^{\rm h}\,39^{\rm m}\,55\fs9$ &  $13^{\rm h}\,40^{\rm m}\,18\fs4$   
                               &  $13^{\rm h}\,37^{\rm m}\,20\fs0$     & $13^{\rm h}\, 41^{\rm m}\, 36\fs7$    & (1) (2) (3) (4) \\
~$\alpha,\delta$(J2000)& --31\degr\,38\arcmin\,24\arcsec  &  --28\degr\,53\arcmin\,32\arcsec 
                                      & --28\degr\,02\arcmin\,42\arcsec  &  --29\degr\,54\arcmin\,47\arcsec   &  \\ 
                                      
                      
~$l,b$                 & 314\fdg86, 30\fdg11  & 315\fdg66, 30\fdg77 & 315\fdg12, 33\fdg74 &    315\fdg72, 31\fdg71  & (1) (2) (3) (4) \\

$v_{\rm opt}$ [\kms]   & $381\pm3$              & $674\pm43$            & $587\pm3$             &   $478\pm3$       & (5) (6) (6) (7) \\ 

type                & Im pec, \HII, Sbrst        & IBm? pec                   & Im                         &   IB(s)m   & (6) (6) (6) (6) \\
\noalign{\smallskip}
distance [ Mpc ]$^a$     & 4.0$^{+0.2}_{-0.4}$           &    4.41         & 4.60       &   4.50  & (8) (9) (9) (9) \\ 
\noalign{\smallskip}
1\arcsec [ pc]       &  19.4        &    21.4          & 22.3       &   21.8            \\
optical diameter    & $5\farcm0 \times 1\farcm9$ & $1\farcm7 \times 1\farcm1$ & $1\farcm3 \times 1\farcm0$ &  $2\farcm5 \times 1\farcm5$  & (6) (7) (6) (7) \\ 
~~~~~~~" [kpc$^2$]  & $5.8\times2.2$                &  $2.2\times1.4$                     & $1.74\times1.4$    &   $3.3\times2.0$ \\

\AB\ [mag] $^b$          & 0.242, 0.70                    & 0.237               & 0.295    &    0.223  & (10,11) (10) (10) (10)   \\ 
$m_{\rm B}$ [mag]   & $10.09\pm0.01$  & $14.73\pm0.21$ & $14.71\pm0.09$ &    $12.36\pm0.09$    & (11) (6) (7) (7) \\    

$m_{\rm H}$ [mag]     & 8.48$\pm$0.02  & $13.9\pm0.2$  &  \nodata  &  $10.85\pm0.07$  &    (11) (TW) (...) (4) \\  

$B-H$               & $1.61\pm0.04$   &   $0.83\pm0.41$   & \nodata       &  $1.51\pm0.16$ &   (11) (TW) (...) (TW) \\
$M_{\rm B}$ [mag]   & $-17.92\pm0.01$ &   $-13.49\pm0.21$       & $-13.60\pm0.09$&         $-15.91\pm0.09$   & (12) (TW) (TW) (TW) \\
\LB\ [10$^8$\Lsun]$^c$ & $22.9\pm0.2$          &   $0.39\pm0.07$  & $0.43\pm0.04$  &     $3.50\pm0.26$   & (12) (TW) (TW) (TW) \\
$M_{star}$ [10$^8$ \Mo]$^d$  &  11.4$\pm$0.5  &  $0.094\pm0.024$ &       0.12$^e$   &   $1.63\pm0.13$   &   (12) (TW) (TW) (TW) \\  
\abox             &    8.18 (A), 8.28 (C) $\pm0.04$ &  8.16$\pm$0.20  &  7.37$\pm$0.20   &  8.50$\pm$0.20  &    (5) (13) (14,TW) (14,TW) \\

\hline
\end{tabular}
\flushleft
$^a$ The errors quoted in NGC~5253 have been estimated taking into account all available measurements (see Sect.~5.2). For the rest of the galaxies, the distance has been obtained from \citet{Karachentsev02}, who didn't quote any uncertainties in their measurements.\\
$^b$ Except for NGC~5253, the \AB\ value tabulated here corresponds to the Galactic extinction \citep*{SFD98}. 
All magnitudes and colors compiled in this Table have been corrected for extinction using the \AB\ value and  following the prescriptions given by \citet{LSE08}.\\
$^c$ The $B$-band luminosity, \LB, is calculated from the extinction-corrected $B$-band absolute magnitude, using a solar $B$-band magnitude of 5.48~mag \citep*{BesselCP98} \\ 
$^d$ The total stellar mass, $M_{star}$,  was computed from the extinction-corrected $H$-band absolute magnitude, assuming a $H$-band mass-to-light ratio of $M_{star}/L_H=0.8$. The $H$-band luminosity, $L_{\rm H}$, is calculated from the $H$-band absolute magnitude, using a solar $H$-band magnitude of 3.35~mag \citep{Colina96}. See \citet{Kirby08} and \citet{LS10} for details.\\
$^e$ For ESO~444-G084 we computed a tentative value of $M_{star}$ assuming $B-H=1.0$ mag.\\

References: 
(1) \citet{TurnerBeck04};  
(2) \citet{Jones05};   
(3) \citet{KaldareCRP03};  
(4) \citet{Skrutskie2MASS06}; 
\mbox{(5) \citet{LSEGRPR07};}  
(6) \citet{deVaucouleurs91};  
(7) \citet{LaubertsValentijn89}; 
(8) \citet{KKHM04};  \\ 
(9) \citet{Karachentsev02}; 
(10)  \citet{SFD98} ; 
(11) \citet{LSE08}; 
(12) \citet{LS10}; \\ 
(13)  \citet*{LeeZG07}; 
(14) \citet*{LeeGH03};  
(TW) This value has been computed in this work.
\end{table*}


Despite all observational effort, the behavior of the neutral gas within NGC~5253 is far from clear. 
The first 
interferometric VLA observations 
\citep{KS95}, 
which only showed the \HI\ map around the galaxy center, revealed that the neutral gas either rotates about the minor axis of the galaxy or flows radially along the minor axis. 
NGC~5253 was observed with the radio-interferometer \emph{Australia Telescope Compact Array} (ATCA) as part of the  \emph{Local Volume HI Survey} (LVHIS)\footnote{http://www.atnf.csiro.au/research/LVHIS} project \citep{Koribalski08,LVHIS}. 
The preliminary LVHIS results  \citep{LSKEGR08} 
showed that the neutral gas is more extended than the optical size of the galaxy, revealing a peculiar morphology in its northern \mbox{areas.} These data also confirmed the intriguing kinematics of the \HI\ emission. Later, \citet{KS08} 
(hereafter, KS08) analyzed high-resolution VLA data of the center of NGC~5253 and found a redshifted \HI\ plume extending to the SE along the minor axis and containing 20--30\% of the neutral gas. These authors speculated that this feature is either a dynamically distinct tidal remnant or an inflow on the far side of the galaxy. They also reported a starburst-blown bubble stalled by interaction with a massive neutral envelope at the NW of the galaxy. However, their data did not allow the analysis of the diffuse neutral gas in which NGC~5253 is embedded, and which provides the key clues to understand all the peculiarities 
observed in this strong starbursting dwarf galaxy.

Here we present full synthesis ATCA \HI\ line data of the BCDG NGC~5253 to analyze its neutral gas distribution in both large and short scales. We also  
perform a comprehensive analysis of this peculiar system combining all available multiwavelength data. 
In addition, we  use the ATCA 
 \HI\ data of the nearby dwarf galaxies ESO~444-G084, IC~4316 and NGC~5264 \citep{LVHIS} 
to compare the properties of NGC~5253 with those found in nearby dwarf galaxies 
within the Centaurus~A group. 
This analysis will give clues to constrain the nature of the starburst phenomenon in NGC~5253. 
Table~\ref{basic} compiles the basic properties of these four galaxies.


The paper is organized as follows: in Section~2 we summarize the observations 
and data reduction; 
Section~3 presents the ATCA \HI\ line results (neutral gas distribution and kinematics) of our analysis of NGC~5253 and the comparison
of the global \HI\ spectrum with the HIPASS data.
Section~4 compiles the 20-cm radio continuum results.
The discussion in Section~5 compares the \HI\ morphology and kinematics (and other galaxy properties, as the star-formation rate, stellar mass, chemical abundances and mass-to-light ratios) of NGC~5253 with those found in similar starbursts reported in the literature and in the nearby dwarf galaxies, as well as analyzes the environment in which NGC~5253 resides.
Section~6 contains our conclusions.

\section{Observations and Data Reduction} 

\HI\ line and 20-cm radio-continuum observations of the galaxies NGC~5253, ESO~444-G084, IC~4316 and NGC~5264
were obtained simultaneously with the \emph{Australia Telescope Compact Array} (ATCA). This radio-interferometer consists of six 22-m antennae, hence giving 15 baselines in a single configuration. 
Our data were obtained between Jan 2005 and Jan 2009  
using three or four different array configurations.
For NGC~5253, we used the
EW367, 750A, 1.5A and 6A
array configurations (the observational details of the other galaxies can be found in Koribalski et al. 2011).
Each galaxy 
was observed for a full synthesis ($\sim$12 hours) in each array to ensure an excellent \emph{uv}-coverage and sensitivity to detect structures 
on all scales. Whereas the short baselines recover the extended \HI\ distribution, the larger baselines provide the details needed to compare
with UV/optical/IR data \citep[e.g.,][]{KLS09}. 
The data in the three most compact array configurations are part of the LVHIS project (Koribalski et al. 2011), while the data in the most extended ATCA configuration are provided by our multiwavelength analysis of a sample of nearby BCDGs 
\citep{LSK09a,LSK09b,LS+11}.

Two different frequency bands are always available at the ATCA. 
We used the first frequency band (IF1) to get the 21-cm line data of neutral hydrogen
(which has a rest frequency of 
1420.406 MHz) and the second frequency band (IF2) to get the 20-cm radio-continuum data. 
IF1 was centered on 1418~MHz with a bandwidth of 8 MHz, divided into 512 channels. 
This gives a channel width of 3.3\kms\ and a velocity resolution of 4\kms. 
The ATCA primary beam is 33\farcm6 at 1418 MHz. IF2 was centered on 1384 MHz (20-cm) 
with a bandwidth of 128 MHz divided into 32 channels. 

Flux calibration was performed by observations of the primary ATCA calibrator, PKS\,1934--638, which has a flux of 14.95~Jy at 1418~MHz. 
The phase calibration was achieved by periodic observations (every 30 minutes) 
of the bright calibration source PKS\,1308--220, which has a flux of 5.04~Jy at 1418~MHz. 
Data reduction was carried out with the {\sc miriad} (\emph{Multichannel Image Reconstruction, Image Analysis and Display}) software package 
\citep*{SaultTW95} 
using standard procedures. After calibration, the IF1 data were split into a narrow 
band 20-cm radio continuum and an \HI\ line data set using a first order fit 
to the line-free channels.

\begin{table*} 
\caption{\label{HIresults} ATCA \HI\ measurements and derived properties. We also list the main properties derived from available single-dish (Parkes) observations, which were compiled in the HIPASS Bright Galaxy Catalogue \citep{Koribalski04} for NGC~5253, ESO~444-G084 and NGC~5264 and in \citet{Minchin03} for IC~4316.}
\centering 
\begin{tabular}{l c c c c}
\hline
Name                    & NGC~5253    & IC~4316   &  ESO~444--G084     &    NGC~5264    \\
HIPASS                  & J1339--31A  & J1340--28 &  J1337--28        &    J1341--29  \\
\hline
\noalign{\smallskip}
{\bf ATCA properties} \\
\noalign{\smallskip}
\HI\ diameter [$\arcmin\times\arcmin$]  & $11\times7$ &  $4\times3$         &   $3\times5$     &      $5.5\times6.5$\\ 
\HI/opt. diameter ratio                  & $\sim2.2\times3.2$   &  $\sim2.4\times2.7$ &       $\sim2.3\times 5.0$       & $\sim2.2\times4.3$    \\

velocity range [\kms]                    & 334 -- 480        & 550 -- 610              &  530 -- 640   &    440 -- 510 \\
$v_{\rm sys}$ [\kms]                    & $404\pm1$       & $579\pm2$            &  $586\pm1$  &    $477\pm1$     \\  
$w_{50}$   [\kms]                         & $63\pm2$         & $22\pm4$              &  $52\pm2$    &   $30\pm2$ \\        
$w_{20}$   [\kms]                         & $95\pm3$       & $40\pm6$              &  $72\pm3$    &   $43\pm3$ \\    
\noalign{\smallskip}   
\HI\ flux density (\FHI) [Jy~\kms]  & $42.1\pm2.3$   & $2.68\pm0.24$      &  $19.9\pm1.3$  &  $8.78\pm0.73$            \\     
\HI\ mass (\MHI) [$10^8$ \Mo]     & $1.59\pm0.09$ & $0.123\pm0.011$  &    $0.992\pm0.065$         &   $0.419\pm0.036$   \\   


\MHI/\LB     [\Mo/\LBo]$^a$               & $0.069\pm0.004$      & $0.313\pm0.034$    &    $2.29\pm0.18$         &    $0.116\pm0.009$ \\
$M_{gas}$/$M_{stars}$ [\%]$^b$     & $18.4\pm1.3$        &  $173\pm46$     &  $\sim1091$        &  $33.9\pm4.0$   \\




\hline

\noalign{\smallskip}
{\bf Single-dish properties} \\
\noalign{\smallskip}

\HI\ flux density (\FHI) [Jy~\kms]                                & $44.0\pm4.5$  & $2.1\pm0.2$ & $18.5\pm2.7$ &   $12.3\pm1.9$  \\

\HI\ mass (\MHI) [$10^8$ \Mo]       & $1.66\pm0.16$ &  $0.096\pm0.009$     & $0.922\pm0.013$       &  $0.59\pm0.09$     \\ 
\MHI/\LB [\Mo/\Lo]$^a$                         &  $0.072\pm0.007$          &  $0.25\pm0.03$               &    $2.12\pm0.32$           &   $0.162\pm0.026$  \\

$M_{gas}$/$M_{stars}$ [\%]$^b$        & 19.2                   & 135      &  1014         &  48   \\
$M_{bar}$ [10$^8$\Msun]$^c$           & $13.6\pm0.7$       &  $ 0.221\pm0.036$  &      $1.34$       & $2.41\pm0.25$       \\


\hline
\end{tabular}
\flushleft
$^a$ The \HI-mass-to-light ratio, \MHI/\LB, is a distance-independent quantity that was computed following \MHI/\LB=$1.5\times10^{-7}$\FHI$\times10^{0.4m_{\rm B}}$, with $m_{\rm B}=m^{NC}_{\rm B}-A_{\rm B}$ \citep*{Warren04}. \\ 
$^b$ The total mass of the neutral gas, $M_{gas}$, was computed assuming $M_{gas}=1.32\times M_{\rm HI}$.\\
$^c$ The baryonic mass, $M_{bar}$, was computed assuming $M_{bar} = M_{stars} + M_{gas}$. 
\end{table*}

By combining data from several array configurations we achieve a very good \emph{uv}-coverage and, 
hence, the Fourier transformation of the obtained visibilities allows us to make datacubes 
and images at a large range of angular resolutions. This is achieved by choosing different weightings for short,
medium and long baselines, which in turn are sensitive to different structure scales. 
The weighting of the data does not only affect the resolution, 
but also the $rms$ noise and sensitivity to diffuse emission.
For our analysis, \HI\ cubes were made using `natural' (na) and `robust' (r=0) 
weighting of the {\em uv}-data in the velocity range covered by 
the \HI\ emission using steps of 4~\kms. 
The longest baselines to the distant antenna six (CA06) were excluded when making the low-resolution  (`na' weighting) and the intermediate-resolution cubes (r=0 weighting), but they were considered to get the high-resolution cube.
Broad-band 20-cm radio continuum images were made using `robust' (r=0) weighting of the IF2 {\em uv}-data and excluding the baselines to CA06. 
The moment maps (integrated intensity map, intensity-weighted mean velocity field, and the velocity dispersion) were created from the \HI\ datacubes by first isolating the regions of significant emission in every channel and afterwards clipping everything below a $\sim2\sigma$ threshold.

\begin{figure*}
\includegraphics[angle=0,width=15.3cm]{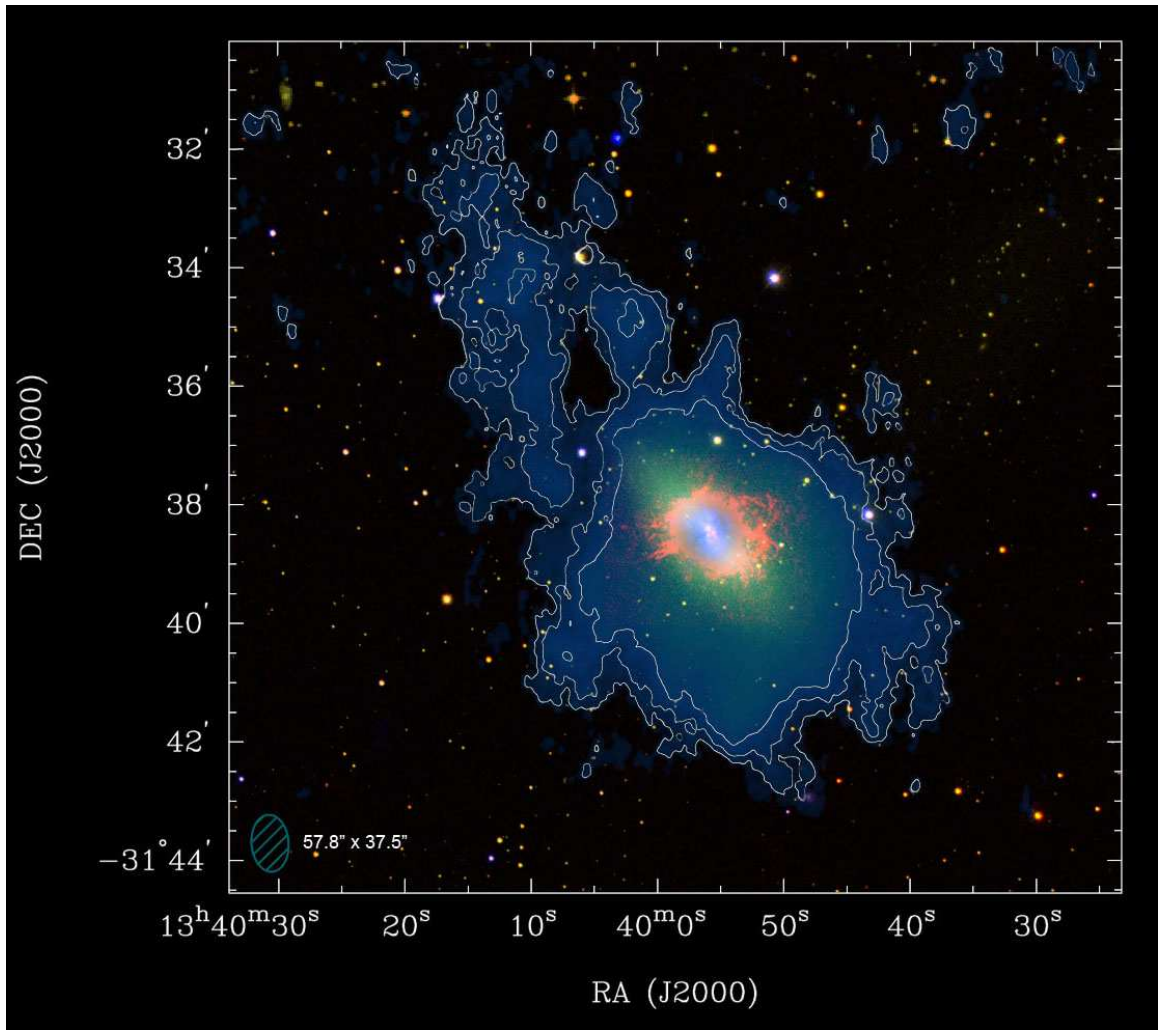} 
\caption{\label{ngc5253color} Color image of the BCDG NGC~5253 obtained combining data in  $FUV$ and $NUV$ (cyan, GALEX), $R$-band (green, Meurer et al. 2006) and \Ha\ (red, Meurer et al. 2006), NIR $J$-band (orange, 2MASS), and our ATCA low-resolution \HI\ map (dark blue).  The synthesised beam of the \HI\ map is $57\farcs8 \times 37\farcs5$.
The plotted contour levels are 0.10 ($\sim3\sigma$), 0.18 and 0.3~Jy\,beam$^{-1}$\kms. 
}
\end{figure*}

\begin{figure*}
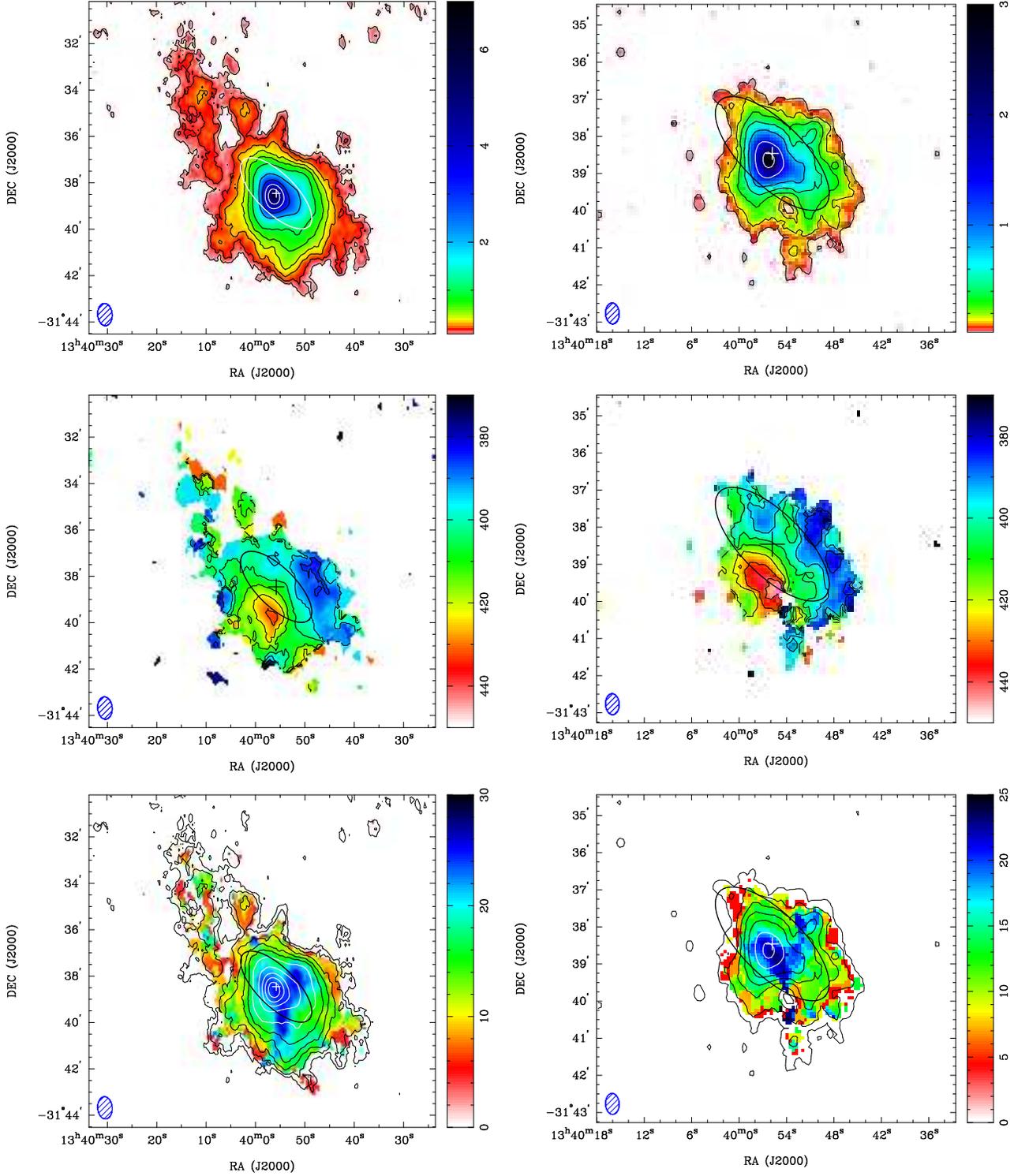

\begin{tabular}{cc}

\includegraphics[angle=-90,width=0.46\linewidth]{ngc5253_mom0.2.FEB11.ps} &
\includegraphics[angle=-90,width=0.47\linewidth]{ngc5253_mom0_rono6.2.FEB11.ps} \\

\includegraphics[angle=-90,width=0.46\linewidth]{ngc5253_mom1.3.ps} &
\includegraphics[angle=-90,width=0.47\linewidth]{ngc5253_mom1.rono6.2.ps} \\

\includegraphics[angle=-90,width=0.46\linewidth]{ngc5253_mom2mom0.2.ps} &
\includegraphics[angle=-90,width=0.47\linewidth]{ngc5253_mom2mom0.rono6.2.ps} \\

\end{tabular}

\caption{\label{ngc5253low} \HI\ moment maps of the galaxy NGC~5253 as obtained from the ATCA, excluding baselines to CA06 and using `natural' weighting (low-resolution datacube, left column) and `robust' ($r=0$) weighting (intermediate-resolution datacube, right column).  The synthesised beam ($57\farcs8 \times 37\farcs5$ and $33\farcs9 \times 22\farcs1$ for the low- and intermediate-resolution datacube, respectively) is displayed in the bottom left corner of each panel. Note that the scale is different for the right and left maps. The galaxy center is always marked with a cross, while an ellipse indicates the size of the optical appearance of the galaxy.
{\bf (top row)} \HI\ distribution (moment 0 map) of NGC~5253, the scale is logarithmic. Contour levels are: 0.10 ($\sim3\sigma$), 0.18, 0.3, 0.45, 0.65, 1, 2, 3, 4 and 5~Jy\,beam$^{-1}$\kms\ (low-resolution datacube) and 0.04 ($\sim3\sigma$), 0.15, 0.3, 0.8, 1.5 and 2.5 ~Jy\,beam$^{-1}$\kms\ (intermediate-resolution datacube). 
    {\bf (middle row)}  Intensity-weighted mean, masked \HI\ velocity field (moment 1 map); the contour levels range from 360  to 460~\kms, in steps of 4\kms.
   {\bf (bottom row)} \HI\ velocity dispersion (moment 2 map), the range is from 0 to 30~\kms\ (low-resolution datacube) and to 25~\kms\ (intermediate-resolution datacube). We overlay the same moment 0 contours we used in the top row maps.}
\end{figure*}








\begin{figure*}

\includegraphics[angle=0,width=15.8cm]{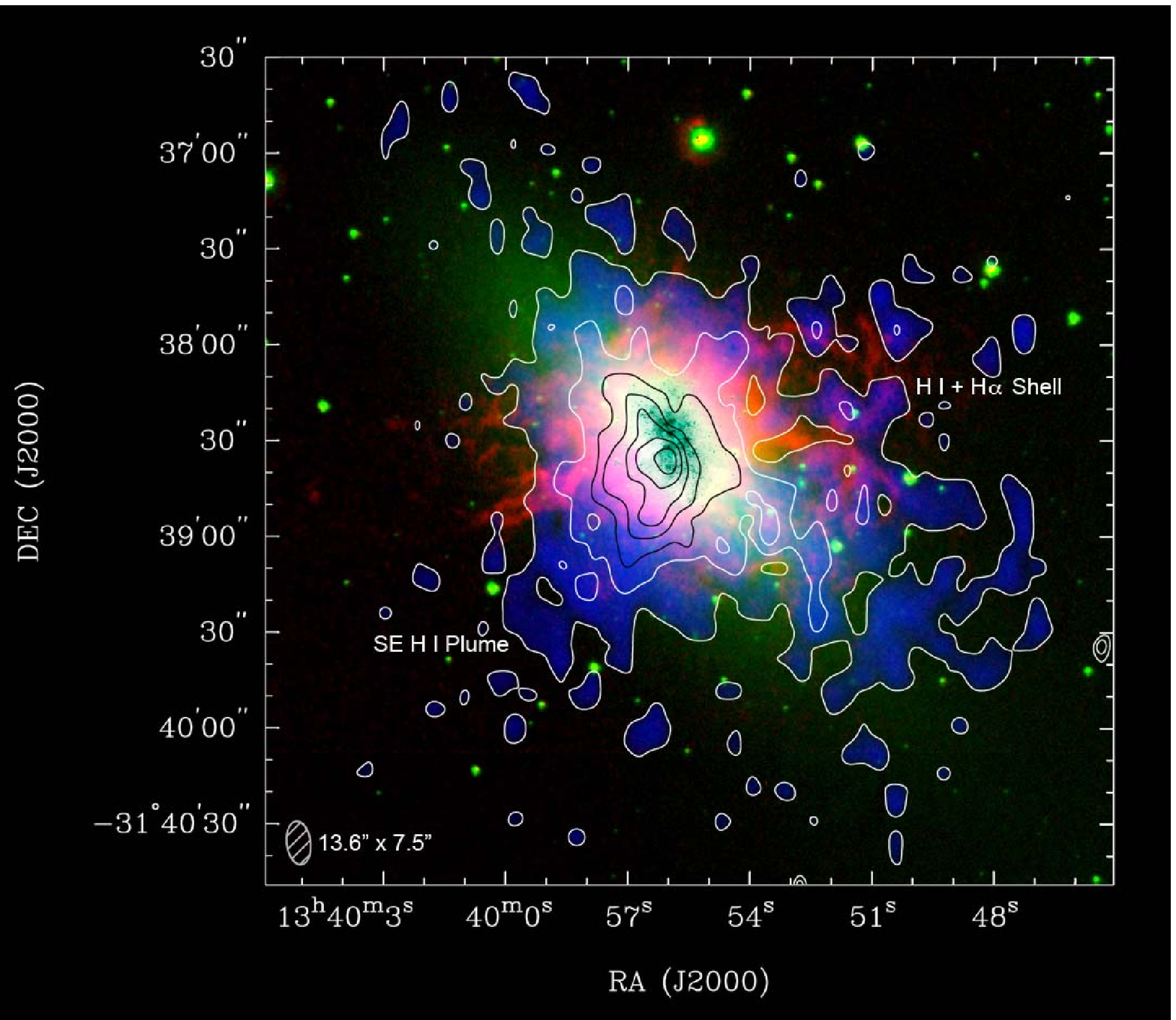}
\caption{\label{ha+hi} Contours of the high-resolution \HI\ map (using `robust', r=0, weighting and including baselines to CA06) overlaid onto a color-image which combines $R$-band and $FUV$-band (green), \Ha\ (red) and the ATCA high-resolution \HI\ map (blue).  The synthesized beam ($13\farcs6 \times 7\farcs5$) is displayed at the bottom left corner. 
The contour levels are 0.032 ($3\sigma$), 0.12, 0.22, 0.32, 0.42, 0.52 and 0.62\,Jy\,beam$^{-1}$\,\kms. 
Note the  \HI\ shell surrounding the NW \Ha\ shell and the \HI\ plume (without any \Ha\ emission associated with it) at the SE, as they were previously noticed by KS08 using VLA data (see their Figure~3).}
\end{figure*}

\section{\HI\ results} 

\subsection{The \HI\ distribution of NGC~5253}


The distribution of atomic neutral gas within the BCDG NGC~5253 is shown in Fig.~\ref{ngc5253color}. This multiwavelength color image compares our new low-resolution \HI\ map (in dark blue) with the ionized gas emission (red) and the stellar component (in green). 
As we see, our new \HI\ map, which is deeper than the previous VLA maps obtained by \citet{KS95} and KS08, clearly shows the diffuse atomic gas emission in the external regions of NGC~5253. 
Indeed, our low-resolution \HI\ map, which is tracing the large scale structures of the neutral gas component of the galaxy, clearly indicates that the atomic hydrogen gas is more extended than the optical size of the galaxy ($\sim 11' \times 6' = 12.8$~kpc~$\times~7$~kpc, that is $\sim2.2\times3.2$ times its optical size), showing a very peculiar morpho\-logy in its northern area. This region shows a kind of filamentary and disturbed \HI\ morphology, which may suggest the existence of tidally disturbed material surrounding the galaxy. The southern region of NGC~5253 also shows a somewhat disturbed morphology, but in this case the \HI\ gas is not as disrupted as it is seen at the northern regions. 




Figure~\ref{ngc5253low} shows the low- and intermediate-resolution \HI\ moment maps of NGC~5253. 
The \HI\ distribution in these maps are shown up to 0.10 and 0.04~Jy~\,beam$^{-1}$\,\kms\ ($\sim3\sigma$ over the noise level), which corresponds to an \HI\ column density of  $5.1\times10^{19}$ and $6.0\times10^{19}$~atoms\,cm$^{-2}$ for the low- and intermediate-resolutions maps, respectively.
The \HI\ maximum coincides with the optical center of the galaxy, although the peak of the \HI\ distribution seems to be displaced to the SE with respect to the optical component of the galaxy. 

The \HI\ emission in our low-resolution datacube covers a velocity range of 330 to 480 \kms. Assuming that the distance to NGC~5253 is 4.0~Mpc (see Table~\ref{basic}) and an integrated flux density of $42.1\pm2.3$~Jy~\kms, we derive a total \HI\ mass of  $(1.59\pm0.09)\times10^8$\,\Mo. The integrated flux in the northern disturbed areas is $\sim$5\% of the total flux of the galaxy, that corresponds to a mass of $\sim8\times10^6$\,\Mo. 

The intermediate-resolution map recovers 81\% of the total flux we detect in our low-resolution map, 
which mainly corresponds to the emission from the main body of the galaxy. 
Indeed, the intermediate-resolution map does not show the diffuse emission at the northern areas, 
although some disturbed material at the southeast can still be observed. 
This region corresponds to the feature that KS08 designed as the `SE \HI\ plume'. We note that this structure is more extended than that reported by these authors, as it extends $\sim3.5\arcmin$ from the center of the galaxy (the low-resolution map presented by KS08 only shows $\sim1.5\arcmin$). Hence, at this resolution the \HI\ distribution in NGC~5253 does not follow its optical major axis but the neutral gas is rather distributed along the minor axis. Furthermore, this distribution is not symmetrical, as the SE region corresponds to the \HI\ plume.  Our intermediate-resolution \HI\ maps also suggest that this feature is not just a plume but appears to be an open-shell structure.  
Most remarkable, the intermediate-resolution \HI\  maps show a hole in the neutral gas emission just at the south of the galaxy and at the west of the `SE \HI\ plume'. This \HI\ hole does not correspond to any of the bubbles that are surrounded by the ionized gas. 



The high-resolution \HI\ map of NGC~5253 is shown in Fig.~\ref{ha+hi}, which displays the \HI\ contours overlaid onto a multiwavelength color picture which combines the \Ha\ image of the galaxy (red) obtained by \citet{Meurer06}, 
the $R$-band and GALEX $FUV$-band images (green) and our high-resolution \HI\ map (blue).
The synthesized beam of our high-resolution \HI\ map is $13\farcs6 \times 7\farcs5$. 
The lowest contour level plotted in  Fig.~\ref{ha+hi} corresponds to 0.032~Jy\,beam$^{-1}$\,\kms\ ($\sim3\sigma$ over the noise level), 
which corresponds to an \HI\ column density of $3.5\times10^{20}$\,atoms\,cm$^{-2}$ assuming that the gas fills the beam. 
Our high-resolution \HI\ map traces the highest density \HI\ gas 
(which corresponds to the small scale structures) 
and agrees quite well with the \HI\ map presented by KS08.
Indeed, both maps have quite similar resolution (their synthesized beam is $9\farcs0 \times 7\farcs6$) and sensitivity (assuming that the gas fill the beam, the lowest \HI\ contour of the high-resolution \HI\ map shown by KS08 corresponds to an \HI\ column density of $6.0\times10^{20}$  atoms\,cm$^{-2}$).  
Our map clearly recovers the three main \HI\ features reported by these authors: 
\begin{enumerate}
\item[1.] the extensions following the optical major axis (in the direction NE--SW and that we clearly identify in the low- and intermediate-resolution maps), 
\item[2.] the `SE \HI\ plume' (that extends beyond the optical component of the galaxy, as it is seen in Fig.~\ref{ha+hi}),  
and 
\item[3.] the open shells at the west and east. 
\end{enumerate}
In particular, Fig.~\ref{ha+hi} shows an excellent correspondence between an \Ha\ shell located at the NE and the \HI\ gas in this region.
Indeed, the \HI\ peak column densities coincide with the \Ha\ peaks in this area.
Furthermore, both the neutral and ionized gas have a hole in the center of this region, 
which also shows high values in its \HI\ velocity dispersion (moment 2 maps in Fig.~\ref{ngc5253low}). 
We derive a neutral hydrogen mass of $\sim5.4\times10^6$ \Mo\ in this \HI\ shell, which matches  
the mass estimation given by KS08. 

On the other hand, we do not see any correspondence between the neutral and the ionized gas 
in the \Ha\ shell located at the east of the galaxy (see Fig.~\ref{ha+hi}).
The only appreciable \HI\ feature in this zone is the `SE \HI\ plume', which lies just at the south of the \Ha\ shell located at the east of the galaxy, 
but this \HI\ plume does not embed any \Ha\ emission.

We also note that Fig.~\ref{ha+hi} shows some residual \HI\ emission following the major axis of the galaxy, specially toward the SW, that was not detected in the maps shown by KS08.

\begin{figure} 
\includegraphics[angle=0,width=8.5cm]{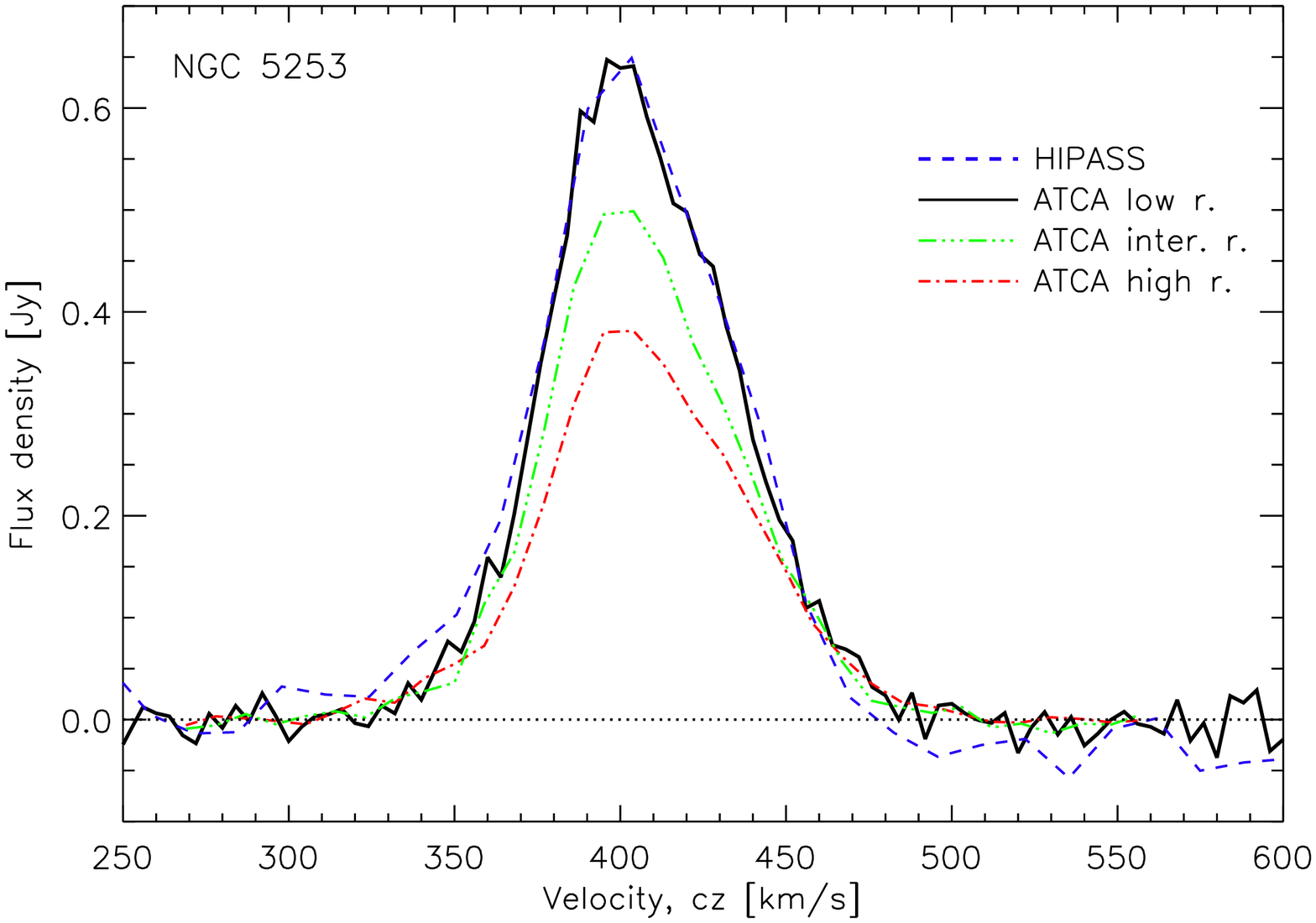} 
\caption{\label{hispectra} The global \HI\  line spectrum 
of NGC 5253 
as obtained from HIPASS (dashed blue line) and LVHIS using the low-resolution datacube (solid black line). The zero baseline is plotted with a dotted black line. 
We also include the global \HI\ line spectra derived using the intermediate-resolution (dashed-dotted red line) and the high-resolution (dashed-dotted green line) datacubes.}
\end{figure}

\begin{figure*}
\includegraphics[angle=0,width=15.7cm]{ngc5253_chan_2_all_MAR11.ps}  
\caption{\label{ngc5253chanlow} \HI\ channel maps, in steps of 9  \kms, of the BCDG NGC~5253 as obtained from 
   the ATCA, using `natural' weighting and excluding baselines to CA06 (low-resolution datacube). 
   The contour levels are -3 (dashed red contours), 3~($\sim2.8\sigma$), 6, 12, 24, 45, 65, and 85 mJy\,beam$^{-1}$. The galaxy center is marked with a red cross, while the green ellipse corresponds in size to the optical appearance of the galaxy.  The center velocity of each channel is displayed 
   at the top left and the synthesised beam ($57\farcs8 \times 37\farcs5$) 
   at the bottom left corner of each panel.}
\end{figure*}

\begin{figure*}
\includegraphics[angle=0,width=15.7cm]{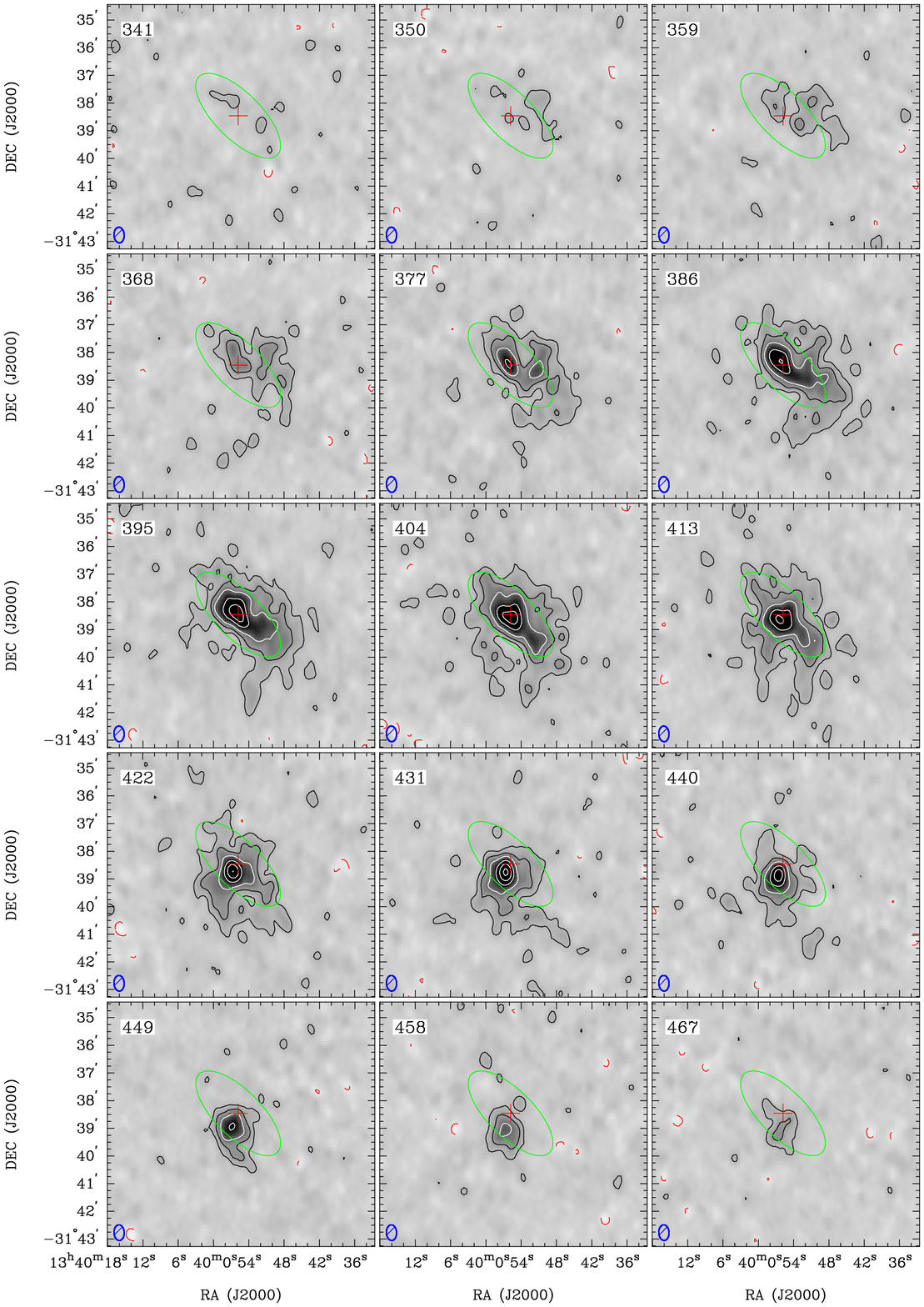} 
\caption{\label{ngc5253chanmed}  \HI\ channel maps, in steps of 9  \kms, of the BCDG NGC~5253 as obtained from 
   the ATCA, using `robust' ($r=0$) weighting and excluding baselines to CA06 (intermediate-resolution datacube). 
   The contour levels are -2.5 (dashed red contours), 2.5 ($\sim2.6\sigma$), 6, 12, 24, 36 and 48 mJy\,beam$^{-1}$. The galaxy center is marked with a red cross, while the green ellipse corresponds in size to the optical appearance of the galaxy.  The center velocity of each channel is displayed 
   at the top left and the synthesised beam ($33\farcs9 \times 22\farcs1$) 
   at the bottom left corner of each panel.}
\end{figure*}

\begin{figure*}
\includegraphics[angle=0,width=14.5cm]{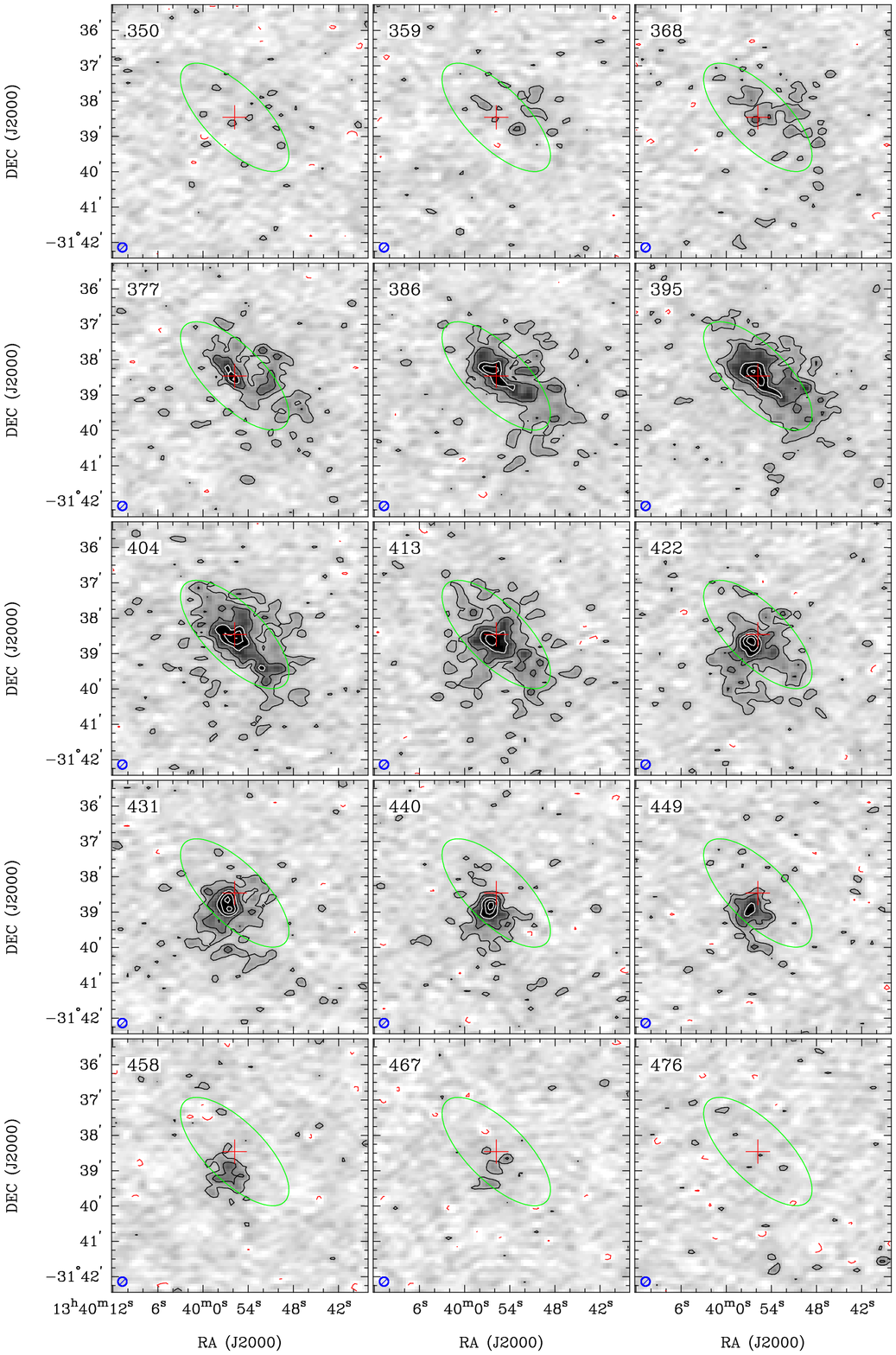}
\caption{\label{ngc5253chanhigh}  \HI\ channel maps, in steps of 9  \kms, of the BCDG NGC~5253 as obtained from the ATCA, using `robust' weighting ($r=0$) and including baselines to CA06 (high-resolution datacube). 
The contour levels are -2.2 (dashed red contours), 2.2 ($\sim2.5\sigma$), 5, 10, 15 and 20~mJy\,beam$^{-1}$. 
The galaxy center is marked with a red cross, while the green ellipse corresponds in size to the optical appearance of the galaxy.  
The center velocity of each channel is displayed   at the top left and the synthesised beam ($15\farcs0 \times 15\farcs0$) 
   at the bottom left corner of each panel.}
\end{figure*}

\subsection{Integrated galaxy spectra and comparison with HIPASS}


Interferometric data usually underestimate the total flux because of the missing short baselines, which filters out any extended diffuse \HI\ emission.
Hence, we compare the integrated spectrum obtained for each galaxy with the single-dish spectrum provided by the
 \emph{\HI\ Parkes All-Sky Survey} \citep[HIPASS; ][]{Barnes01, Koribalski04}. 
The HIPASS spectrum of each galaxy was extracted from the raw HIPASS datacubes assuming a point-source (their intrinsic size is smaller than the $\sim16\arcmin$ angular resolution of the gridded data) within a region of $20\arcmin\times20\arcmin$ centered on the source, with each pixel weighted according to the expected beam response. Following the procedure described in \citet{Koribalski04} 
and revisited by \citet{Kirby10}, 
we derived the basic spectral properties of the galaxies, which for NGC~5253, ESO~444-G084 and NGC~5264
are virtually identical to those compiled in the HIPASS Bright Galaxy Catalogue \citep{Koribalski04}.
Similar single-dish values for integrated \HI\ properties of these three galaxies are provided by the HIDEEP Survey \citep{Minchin03}. 

Figure~\ref{hispectra} shows a comparison between the global \HI\ line spectrum as obtained from HIPASS (dashed blue line) and LVHIS (solid black line) for NGC~5253. The agreement is quite good, LVHIS data recover almost 96\% of the single-dish \HI-flux, suggesting that very little diffuse \HI\ emission has been filtered out in our interferometric observations. 

\subsection{The \HI\ kinematics of NGC~5253}

The central panels in Fig.~\ref{ngc5253low} show the mean, masked low- and intermediate-resolution \HI\ velocity field of the galaxy. 
As previously noticed by \citet{KS95}, the most remarkable characteristic of these maps is that the neutral gas \emph{is not} rotating about the optical minor axis of the galaxy (as it should be expected) but \emph{it seems to rotate about its optical major axis}, as it happens in polar ring galaxies, or it just flows radially along the minor axis. Later, KS08 used high-resolution maps to discard the rotation scenario, emphasizing that their new data are more consistent with either an inflow or outflow of neutral gas along the minor axis from the south and east of the nucleus. 

Here we will consider both scenarios to try to explain the intriguing and disturbed \HI\ kinematics of NGC~5253.
Figures~\ref{ngc5253chanlow}, \ref{ngc5253chanmed} and \ref{ngc5253chanhigh}  show, respectively, \mbox{the low-,} \mbox{intermediate-, and} high-resolution \HI\ channel maps of NGC~5253. 
As we see, the \HI\ distribution is asymmetric and does not show any clear rotational pattern. 
To facilitate the analysis of the \HI\ kinematics within this peculiar BCDG, we plot in Figs.~\ref{ngc5253pv} and \ref{ngc5253pv2} some position-velocity (pv) diagrams across our \HI\ datacubes. The position angle (PA) and localization of these slits are shown in Fig.~\ref{ngc5253slit}.  
The results found in our analysis are the following:

\begin{figure}
\includegraphics[angle=0,width=\linewidth]{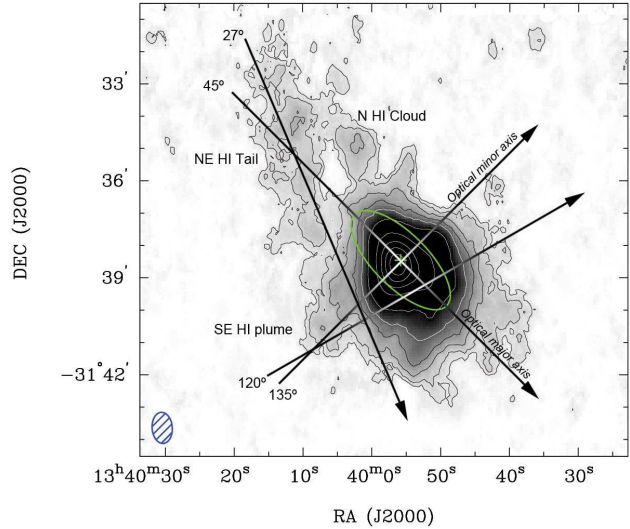}
\caption{\label{ngc5253slit} \HI\ distribution map of NGC~5253 using our low-resolution datacube and showing the slits positions used to create the position-velocity diagrams shown in Figs.~\ref{ngc5253pv} and \ref{ngc5253pv2}. The most prominent features have been labeled.}
\end{figure}

\begin{figure}
\begin{tabular}{c}
\includegraphics[angle=-90,width=0.92\linewidth]{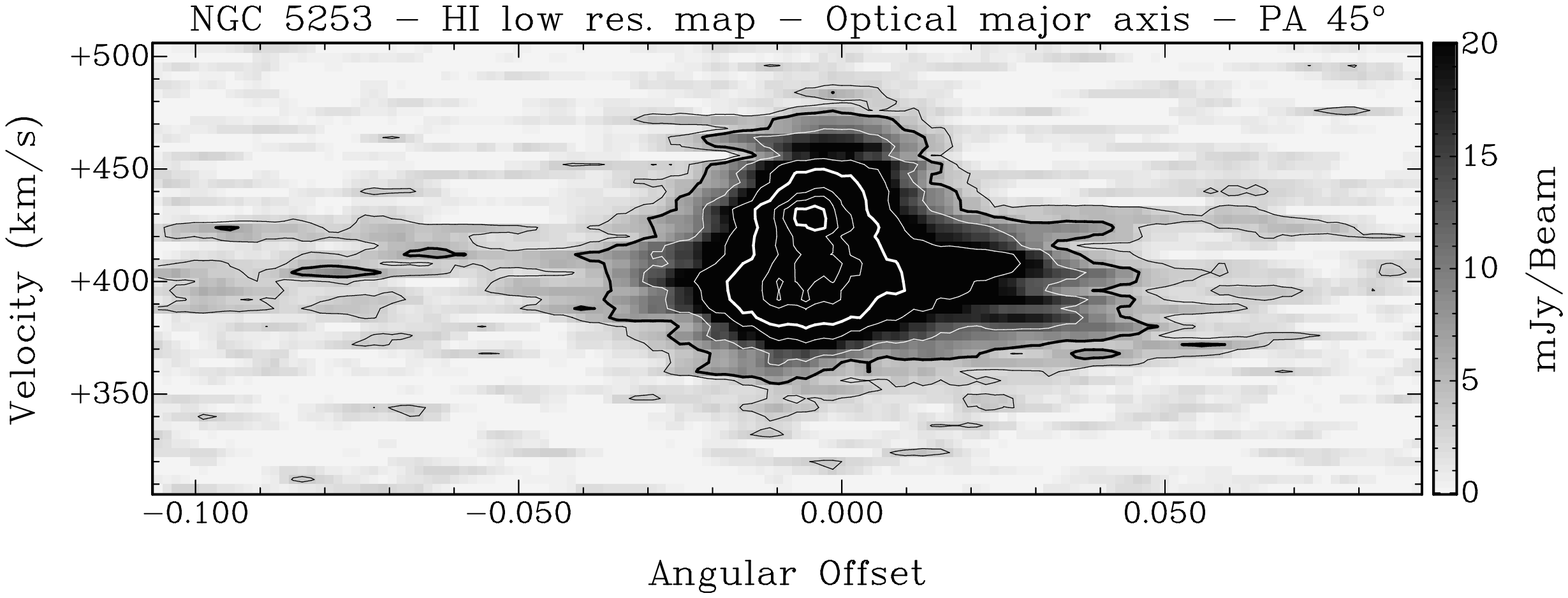} \\
\includegraphics[angle=-90,width=0.92\linewidth]{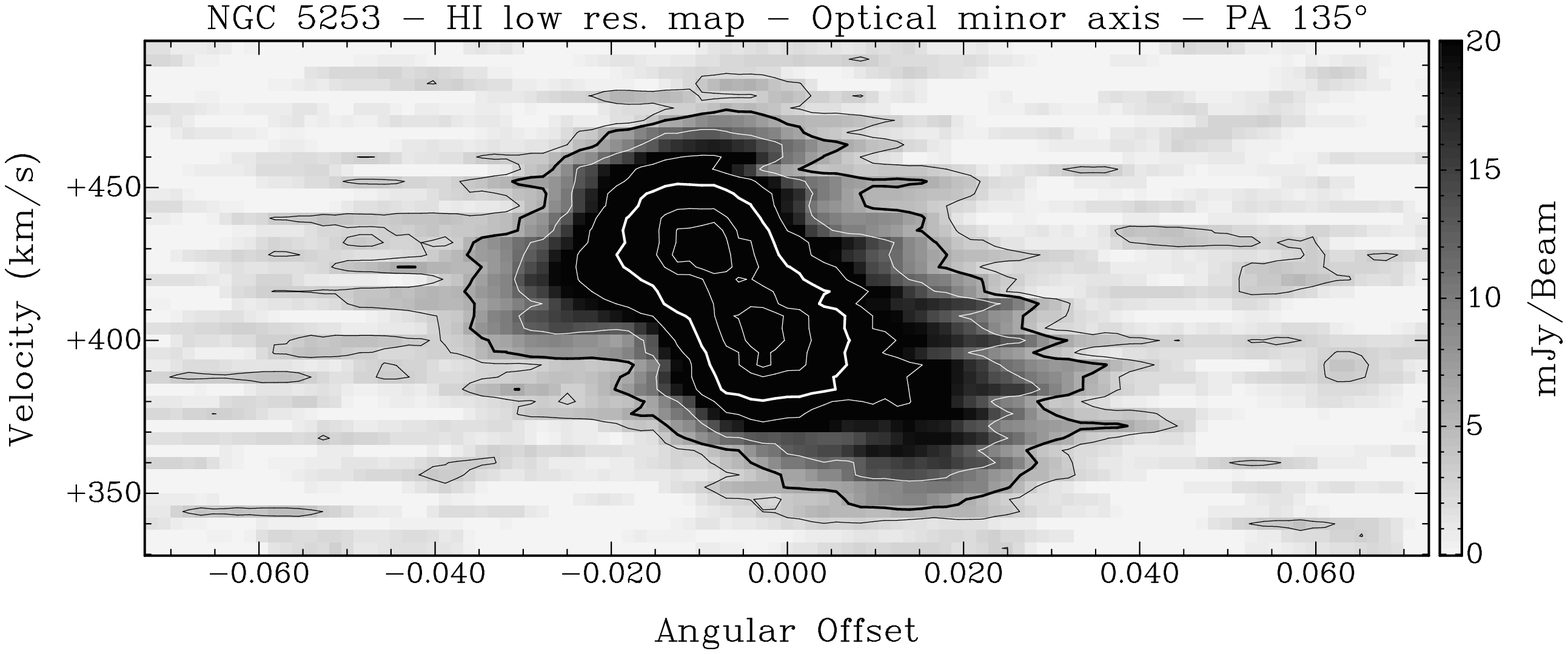}\\
\end{tabular}
\caption{\label{ngc5253pv} \HI\ position-velocity diagrams of NGC~5253 using our low-resolution datacube. The slits cross along the optical major axis (top panel, with PA=45$^{\circ}$, northeast is left) and along the optical minor axis (bottom panel, with PA=135$^{\circ}$, southeast is left). 
The units of the $x$-axis are radians, 0.1 radians are 5.73$^{\circ}$ or 400~kpc at the distance of NGC~5253.
The  contour levels are: 3, 6, 12, 25, 40, 60, 70 and 80~mJy\,beam$^{-1}$. The slit positions are shown in Fig.~\ref{ngc5253slit}.}
\end{figure}

\begin{enumerate}
\item[1.] The pv diagram along the optical major axis of NGC~5253 (with a PA of 45$^{\circ}$) is plotted in the top panel of Fig.~\ref{ngc5253pv}. The neutral gas seems to show a constant velocity along the slit, with a velocity almost coincident with the systemic velocity ($\sim410$~\kms). 
However, the central regions show an elongation to higher velocities 
($\sim$470~\kms). A not so obvious elongation to lower velocities ($\sim350$~\kms) is also observed in the same region. 
Indeed, the channel maps 
almost always show \HI\ emission in the center of the galaxy. 
However, the velocity minimum 
is located at $\sim1.25$~arcmin from the position of the velocity maximum. This feature seems to be a reminiscent of a rotation pattern which follows the optical major axis of the galaxy. KS08 already suggested the presence of such feature in their high-resolution \HI\ maps (see their figure~9), but our pv diagram 
(which shows \HI\ emission $\sim10'$ along the optical major axis)
shows more clearly this behavior. 
We also note that there are two maxima in the \HI\ emission just in the very center of the galaxy, that may suggest the existence of two main entities, as we will discuss below. 

\item[2.] The pv diagram along the optical minor axis of NGC~5253 (with a PA of 135$^{\circ}$, which is plotted in the bottom panel of Fig.~\ref{ngc5253pv}) 
shows an evident velocity gradient of the neutral gas, from $\sim470$~\kms\ at the east to $\sim350$~\kms\ at the west. That is the main velocity field that we obtain in the first moment maps (see Fig.~\ref{ngc5253low}). We also detect two \HI\ peaks in this pv diagram 
embedded in two structures that have different orientations.  
From the western peak towards the west, the velocity decrease continuously from $\sim400$~\kms\ to $\sim350$~\kms. However, the eastern region (which corresponds to the `SE \HI\ plume') shows a decrease of the velocity from the eastern \HI\ peak ($\sim440$~\kms) to the east ($\sim400$~\kms). 
The observed overall sinusoidal pattern, which is characteristic of interactions between two objects, is also found in the intermediate- and high-resolution datacubes.
\item[3.] The intermediate-resolution  \HI\ channel maps of NGC~5253 show two open bubbles or shells at the NW (between 359 and 386~\kms) and at the SE (between 404 and 422~\kms). These features may correspond to outflows of material originated in the central starburst. 
We analyze individually these two features:

\begin{figure}
\begin{tabular}{c}
\includegraphics[angle=-90,width=0.92\linewidth]{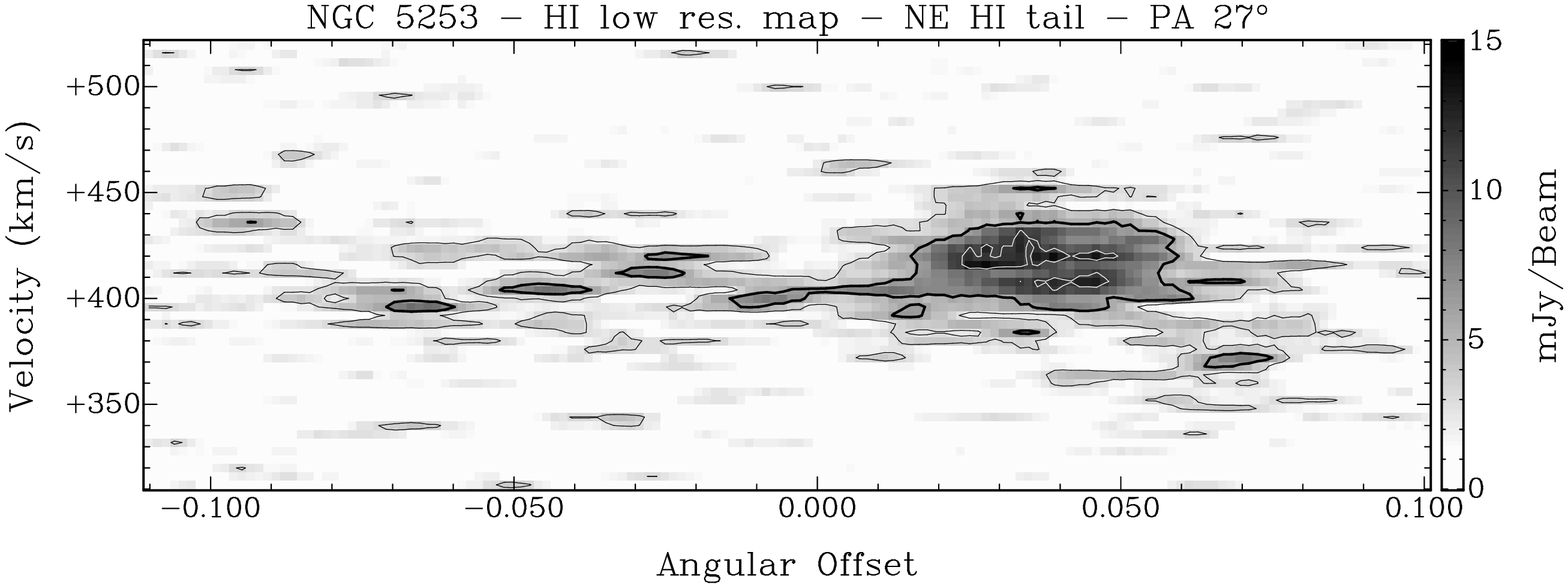} \\
\includegraphics[angle=-90,width=0.92\linewidth]{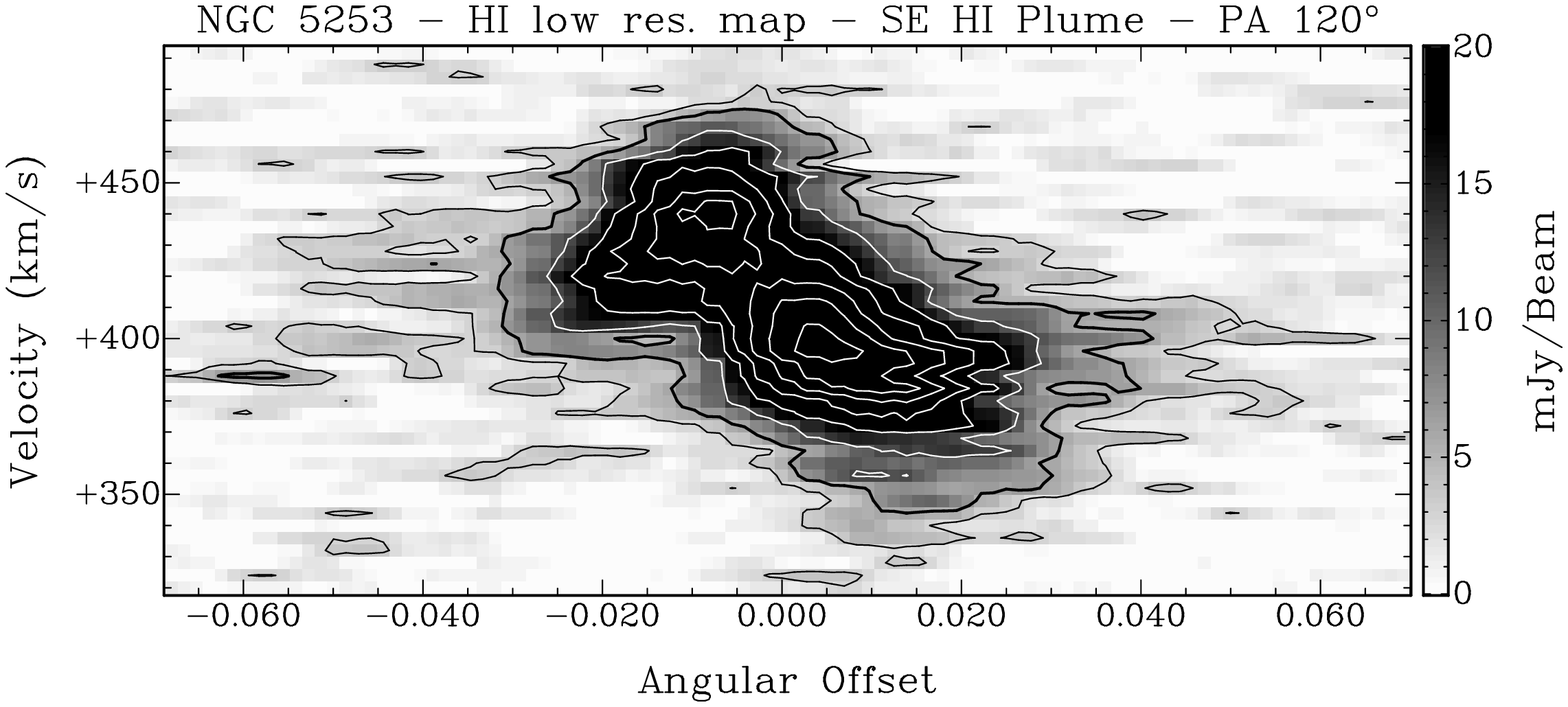} \\
\end{tabular}
\caption{\label{ngc5253pv2} \HI\ position-velocity diagrams of NGC~5253 using our low-resolution datacube. The slits cross along the disturbed material at the NE of the galaxy (top panel, with PA=27$^{\circ}$, northeast is left) and along the `SE \HI\ plume' (bottom panel, with PA=120$^{\circ}$, southeast is left). 
The units of the $x$-axis are radians, 0.1 radians are 5.73$^{\circ}$ or 400~kpc at the distance of NGC~5253.
The  contour levels are: 3, 6, 12, 24, 31, 38, and 44~mJy\,beam$^{-1}$. The slit positions are shown in Fig.~\ref{ngc5253slit}.}
\end{figure}

\begin{enumerate} 
\item[(a)] The NW shell corresponds to a bubble of neutral and ionized gas (`\HI+\Ha\ Shell' in Fig~\ref{ha+hi}) originated in the densest regions of the ISM. 
Indeed, this structure, which has a size of $\sim0.8-1$\,kpc,
is traced by fragmented \HI\ gas in the 377--413~\kms\ channel maps of the high-resolution \HI\ datacube (Fig.~\ref{ngc5253chanhigh}).
The `NW \HI+\Ha\ Shell' was already detected by KS08, who noticed that
the \HI\ velocity range (30--40~\kms) is comparable to that found in the ionized gas. 
The non-detection of X-ray emission 
indicates that it has not been produced by the most recent star-formation event, 
but it was originated $\sim$40~\Myr\ ago. 
\item[(b)] The `SE \HI\ shell' (between 404 and 422~\kms\ in Fig.~\ref{ngc5253chanmed}) 
has lower \HI\ column densities than those seen in the NW shell.
The eastern areas actually correspond to the `SE \HI\ plume' reported by KS08 and has no counterpart to any \Ha\ or X-ray emission. 
This shell has a very large size, $\sim1.8-2$~kpc, i.e., it is twice as large as the `NW \HI+\Ha\ shell', and as large as the optical minor axis of the galaxy. 
We do not consider that this structure has been originated by an outflow.
A careful analysis of the previous (368--395\kms) and later (431--449~\kms) channel maps indicate a continuous evolution of the \HI\ gas from the NW to the SE (along the optical minor axis, the SE region corresponds to the `SE \HI\ plume') and from the NE to the SW (along the optical major axis), that may correspond to two different entities.
\end{enumerate}

\item[4.] The disturbed northern region observed in the low-resolution \HI\ distribution map is composed by several small \HI\ clouds, which have a relative large range of velocities  (386 to 458~\kms). We identify two main structures which do not show any stellar counterpart and are not related to any \Ha\ emission, suggesting that both have a tidal origin:

\begin{enumerate}
\item[(a)] The brightest of the \HI\ clouds is located just at the north of the NE end of the stellar distribution of the galaxy.
This `N \HI\ cloud' has a total \HI\ mass of \mbox{$\sim2.0\times10^6$~\Mo}
and it is connected to the main body of the galaxy by diffuse neutral gas. 
\item[(b)] The elongated and narrow \HI\ tail at the NE of the system 
consists of a chain of several \HI\ clumps. Although they have a relatively constant velocity of around 400~\kms, the clumps also show some variations in velocity following a kind of sinusoidal pattern with an amplitude of $\sim$20--30~\kms\ (see top panel of  Fig.~\ref{ngc5253pv2}).
The total \HI\ mass in the `NE \HI\ tail' is $\sim4.2\times10^6$~\Mo. 
It is very interesting to note that there is a smooth transition in both position and velocity between the eastern tip of the SE \HI\ plume (which has a velocity of $\sim$400~\kms) and the `NE \HI\ tail' (which has a similar radial velocity in this area). 
\end{enumerate}

\begin{figure*}
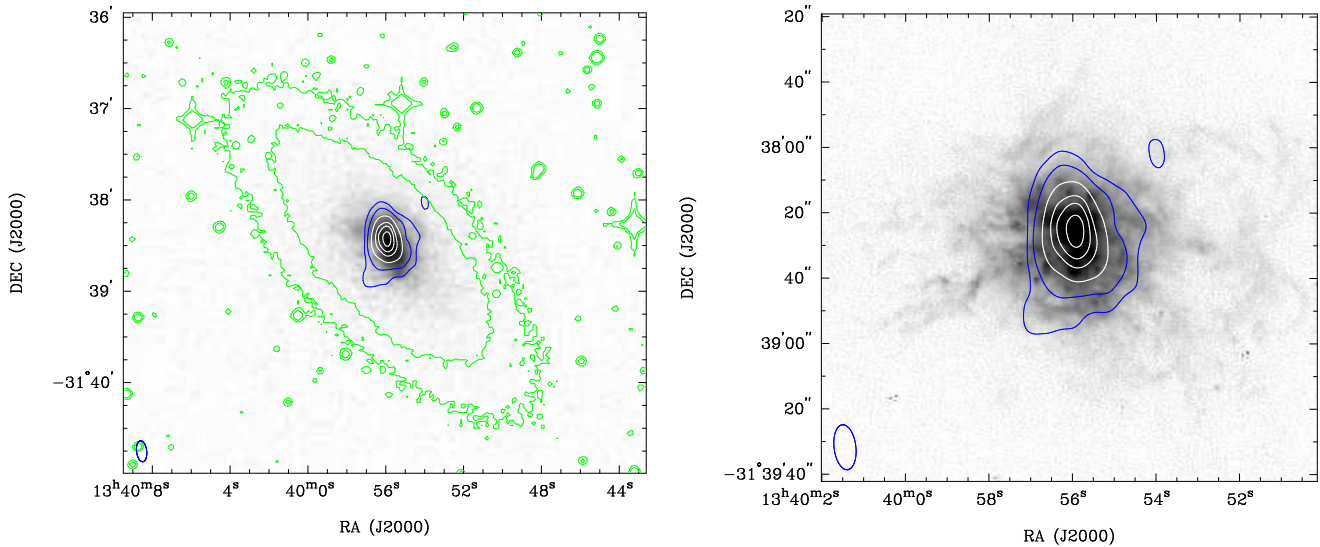

\centering
\begin{tabular}{cc}
\includegraphics[angle=-90,width=8.5cm]{NGC5253_20cm+FUV.ps} &
\includegraphics[angle=-90,width=8.5cm]{NGC5253_20cm+Ha.ps} \\
\end{tabular}
\caption{\label{ngc5253rc} 20-cm radio-continuum contours of NGC~5253 overlapped onto a {\bf (left)} GALEX $FUV$-band image and {\bf (right)} a \Ha\ image (Meurer et al. 2006), both images on logarithmic scale. The contours are  -0.9, 0.9, 1.8, 4, 9, 14 and 22 mJy\,beam$^{-1}$. The left panel shows the contours of the $R$-band image in green.
The synthesized beam ($13\farcs8 \times 6\farcs7$)  is displayed at the bottom left corner in both panels. }
\end{figure*}

\item[5.] Finally, some channel maps show long plumes of neutral gas extending from the \HI\ maxima towards the external regions of the galaxy. Specially evident are the features seen in the 368--386~\kms\ and in the 431--440~\kms\ channel maps (the later structure corresponds to the `SE \HI\ plume'),
which are both spatially coincident at the southern regions of NGC~5253. 
Indeed, the pv diagram crossing this region (bottom panel of Fig.~\ref{ngc5253pv2}) clearly shows 
two independent \HI\ peaks embedded in a sinusoidal pattern.
The `SE \HI\ plume' increases velocity
from the east ($\sim400$~\kms) to the center of NGC~5253 ($\sim460$~\kms), but then this tendency is broken, as the velocity of the SW region decreases from the center of the galaxy ($\sim410$~\kms) to the external western areas ($\sim350$~\kms). 
We notice that if we consider both structures as a single entity the global kinematical pattern is a decrease of the velocities from the east ($\sim460$~\kms) to the west ($\sim350$~\kms), that could be incorrectly attributed to rotation. As we discussed before, the `SE \HI\ plume' is not associated with any of the structures observed in the \Ha\ images, and hence its outflow origin can be discarded (see also KS08). Therefore, the most plausible explanation of the sinusoidal pattern is that we are observing two different entities 
that are now in interaction.
The 3D analysis of the low- and intermediate-resolution datacubes 
indicates a continuity between the features located at the NW regions 
(as the western \HI\ peak in the 368~\kms\ channel map) 
with the `SE \HI\ plume' following the minor axis, which continues to the diffuse tidal material at the NE. This complete entity seems to be an independent system which is falling into NGC~5253. The main body of the galaxy would then correspond to the SW and NE areas, which are aligned with the optical major axis and still is rotating.
\end{enumerate}


In summary, these new, deep \HI\ data seem to indicate that the neutral gas found in 
the northern areas of the galaxy have a tidal origin. Hence,
we propose that the very peculiar \HI\ morphology and kinematics of NGC~5253 can be explained by 
an interaction scenario, in which the infall of an independent 
\HI\ cloud along the minor axis of the galaxy is disturbing the existing neutral gas and triggering the powerful starburst. 
Both entities now seem to be in a process of merging, but the interaction probably started a long time ago,
as we detect tidal gas at the northern regions of the galaxy and it seems that previous starbursts 
have occurred in the system.
We also confirm the existence of a high-density \HI\ bubble at the NW regions, which matches the \Ha\ emission and 
seems to be a consequence of the combined actions of the winds of massive stars and supernova explosions. 
However, our \HI\ data alone do not allow to definitively confirm the proposed infall scenario, as some features (the long disturbed neutral gas at the NE  and the compressed HI edge at the SW) may have also been originated as the result of ram pressure stripping from moving through a dense IGM.


\section{20-cm Radio-continuum emission} 



Figure~\ref{ngc5253rc}  shows the contours of the 20-cm radio continuum emission map of NGC~5253 
and its comparison with the GALEX $FUV$-band and \Ha\ images. 
The radio-continuum emission within this BCDG is clearly resolved 
and has an extension of $0.6\arcmin\times1.0\arcmin$. 
The 20-cm radio-continuum emission is located exactly at the center of the galaxy, at the position where the powerful starburst 
(and almost all the $FUV$ emission) is found, and shows an extension towards the SE. 
This faint feature may be related to an \Ha\ arc that is located in this area.
 We also detect faint emission at the NW of the galaxy, which is detached from the main 20-cm emission of NGC~5253. This object, which has an 1.4~GHz flux of $1.2\pm0.2$~mJy, is located over some faint \Ha\ emission in the external region of the galaxy. However, it is probably a background radio-source.

The total 20-cm radio continuum flux we derive from NGC~5253, $S_{\rm 1.4\,GHz}=78\pm3$~mJy, is slightly lower than the value derived by \citet{Condon+96}
in their 1.4~GHz VLA Atlas of bright IRAS galaxies, which was 83.8~mJy, although the synthesized beam size of their 20-cm radio-continuum map was larger ($54\arcsec$). Assuming that the galaxy lies at a distance of 4.0~Mpc and applying the expression given by \citet*{YRC01}, we derive a 1.4~GHz luminosity of $(1.47\pm0.06)\times 10^{20}$~W\,Hz$^{-1}$ for NGC~5253.


\section{Discussion} 

\begin{table*} 
\caption{Luminosities and star formation rates [\MMoy] of NGC~5253, ESO~444-G084, IC~4316 and NGC~5264.}
\label{sfr} 
\begin{tabular}{l c c c c}
\hline
Name                    & NGC~5253    & IC~4316   &  ESO~444--G084     &    NGC~5264  \\
HIPASS                  & J1339--31A  & J1340--28 &  J1337--28        &    J1341--29  \\
\hline

$L_{\rm FUV}$ [$10^{38}$ erg\,s$^{-1}$ \AA$^{-1}$]
	             & $23.5 \pm 1.5$    & $1.54 \pm 0.11$               &   $1.15 \pm 0.14$          &    $2.18 \pm 0.30$        \\
$L_{\rm H\alpha}$ [10$^{38}$ erg\,s$^{-1}$]        
                    & $440 \pm 20$     &     $ \sim1.90$                  &    $4.10\pm0.38$           &         $16.4\pm2.9$      \\  
$L_{\rm 60\mu m}$ [10$^{22}$ W\,Hz$^{-1}$]  
	             & $5.71 \pm 0.01$   & \nodata                &  \nodata           &         $0.0737\pm0.0013$       \\
$L_{\rm FIR}$ [10$^{41}$ erg\,s$^{-1}$]             
                     & $25.83 \pm 0.09$    & \nodata                  &   \nodata         &    $0.475\pm0.090$              \\
$L_{\rm 20cm}$ [10$^{19}$ W\,Hz$^{-1}$]  
                    & $14.7 \pm 0.6$  &    $0.126\pm0.039$                      &           $ < 0.6$ &    $< 0.7$           \\
\hline
\sfrfuv\ (S07)      & $0.190 \pm 0.010$   & $0.0125 \pm 0.0009$               &   $0.0093 \pm 0.0011$          &   $0.0177 \pm 0.0025$       \\ 
\sfrha\ (C07)       & $0.233 \pm 0.011$    & $\sim0.0010$     &       $0.0022\pm0.0002$      &    $0.0087\pm0.0015$        \\
$SFR_{60\mu m}$     & $0.11$     & \nodata               &   \nodata            &     $0.0014\pm0.0003$          \\
\sfrfir\            & $0.12 $   & \nodata             & \nodata               &     $0.0021\pm0.0004$           \\
\sfrghz 
                    & $0.176 \pm 0.010$   & $0.0015 \pm 0.0005$   &    $<0.0071$     &  $<0.0086$ \\

 \noalign{\smallskip}

Assumed $SFR$ &   0.20  &  0.0035 &   0.0070   & 0.0090    \\
 
\hline
\end{tabular}
\flushleft
References: 
   S07 = \citet{Salim07} ;
   C07 = \citet{Calzetti07}. 
\end{table*}

\subsection{The global star-formation rate of NGC~5253 and nearby dwarf galaxies}

The star formation rate of a galaxy can be estimated using different calibrations at many different wavelengths, from X-ray to radio. 
Here, we will consider the global SFR derived from (i) far-ultraviolet ($FUV$) images using the calibration provided by \citet{Salim07}, (ii) \Ha\ images and the calibration given by \citet{Calzetti07}, (iii) far-infrared ($FIR$) data, in particular, the 60~$\mu$m flux with the calibration provided by \citet{Condon92} and the total $FIR$ flux with the calibration given by \citet{K98}, and (iv) 20~cm radio-continuum data using the calibration provided by \citet*{CCB02}. 
A recent review of all these calibrations is provided by \citet{LS10}, 
who analyzed the $SFRs$ of a sample of strong starbursting systems (including NGC~5253) using multiwavelength data. 
For comparison purposes, we also computed the $SFRs$ derived for the other dwarf galaxies of the M~83 subgroup.
Table~\ref{sfr} lists both the luminosities and the SFRs derived for these four galaxies using this multiwavelength analysis.


We note that the $FUV$ and \Ha\ luminosities quoted in Table~\ref{sfr} have been corrected for extinction in all galaxies using the $A_B$ values indicated in Table~\ref{basic} and following the prescriptions explained in \citet{LSE08} and \citet{LS10} for correcting the \Ha\ and the $FUV$ fluxes, respectively. The total \Ha\ fluxes of the galaxies NGC~5253, ESO~444-G084 and NGC~5264 have been extracted from \citet{Meurer06}. To estimate the total \Ha\ flux in IC~4316, we added the \Ha\ flux of the six star-forming regions spectroscopically observed by \citet*{LeeZG07} 
within this galaxy. The value was corrected for reddening considering an average value for the $C_{\rm H\beta}$ coefficient these authors estimated for the star-forming regions, for which we derive $A_B\sim0.63$ mag. However, the total \Ha\ flux computes in this way underestimates the total \Ha\ flux of IC~4316 because the spectroscopic data compiled by \citet{LeeZG07} do not cover all the ionized gas detected in the galaxy. Considering the aperture sizes and the \Ha\ map provided by these authors (see their Fig.~2b) we estimate that the area recovered by the spectroscopic data is $\sim20$\% of all the region showing \Ha\ emission. The values compiled in Table~\ref{sfr} consider this correction and, therefore, are only tentative estimations to the real \Ha\ values.
%

For NGC~5253, the \Ha-based and the $FUV$-based SFRs agree quite well.
The $FIR$-based $SFR$s also are in agreement with those values, however, these SFR estimations have a larger uncertainty. 
Interestingly, the \sfrghz\ value also agrees with them.
Note that \Ha\ emission traces the most recent star-formation event ($<10$~Myr) but $FUV$ and radio-continuum emission trace the star-formation activity in the last $\sim$100~Myr. Hence, although the most recent star-formation event happened $\sim3.5$~Myr ago \citep{LSEGRPR07} and 
 the majority of the star clusters located in the center of the galaxy have ages younger than 10~Myr \citep{Alonso-Herrero04}, the  good agreement between all multiwavelength SFR indicators suggests that the elevated star formation activity has been present in NCG~5253 for several $\sim$100~Myr. This result agrees very well with the recent analysis presented by \citet{McQuinn+10a,McQuinn+10b}, who measured the duration of starbursts in twenty nearby, ongoing, and {\it fossil} starbursts in dwarf galaxies, including NGC~5253. This analysis, that was based on the recent star formation histories derived from resolved stellar population data obtained with the {\it Hubble Space Telescope}, revealed that the starburst durations may range from 450 to 650~Myr (sometimes even $\sim$1.3~Gyr). 


On the other hand, for the other galaxies the $FUV$-based $SFR$ is always higher than the \Ha-based SFR, with both quantities also higher than the \sfrghz. 
This suggests that the recent star-formation activity has decreased when compared with that experienced by the galaxies in the last $\sim$100~\Myr. We note, however, that some recent analyses \citep*[i.e.,][]{Lee09,Pflamm-AltenburgWK09}
indicate that the \Ha\ luminosity underestimate the $SFR$ relative to the $FUV$ luminosity  in dwarf galaxies with $SFR\le0.01$~\Moy. But the very low 20-cm radio-continuum flux estimated for these three galaxies strongly suggests that the star-formation activity has not been strong recently. 


\subsection{Comparison of the neutral gas component of NGC~5253 with that found in other starbursts}

We now compare the features of the neutral gas component found in NGC~5253
with the other 
starbursts described in the literature. 
%
%
%
Recent analyses of the neutral gas component of dwarf starburst galaxies include NGC~2366 \citep{vanEymeren09a},
NGC~4861 \citep{vanEymeren09b}, NGC~5408 and IC~4662 \citep{vanEymeren10}. All these galaxies have the common characteristic of showing outflows of neutral gas that follows the ionized gas. Indeed, the minimum of the \HI\ column density are usually find at the center of these structures, as it happens in the NW region in NGC~5253 (see Fig.~\ref{ha+hi}) where, as previously noticed by KS08, and \HI\ shell spatially coincides with an \Ha\ shell, and both structures seem to have similar expansion velocities (30--40~\kms). 
This effect seems to be a consequence of both (i) the consumption of the neutral gas to form new stars and (ii) the injection of kinetic energy into the ISM by the combined action of both winds from massive stars and supernova explosions, that pushes the ionized gas and creates a superbubble that may develop in a galactic wind. However, perhaps the large amount of neutral gas still found in these shells are responsible for stalling the expansion and prevents the formation of a galactic wind (KS08), and hence that is the reason why these are not found in dwarf galaxies. 
Therefore, the non-detections of galactic winds in dwarf starburst galaxies will contradict the theoretical models by \citet{MacLowFerrara99} but will agree with the theoretical models developed by \citet{SilichTT98}. Following these models, the expansion velocity of the outflows will never get close to or beyond the escape velocity due to presence of a dark matter halo (in our case, maybe the huge amount of neutral gas that is surrounding the main body of the galaxy is also playing an important role) that slows down the gas expelled by the galactic winds. In fact, the expansion velocities of the gas computed by \citet{vanEymeren10} are 30-50\% the escape velocities.

A clear \HI\ blowout is found in the dIrr galaxy Holmberg~I (Ho~I), belonging to the M~81 galaxy group \citep{Ott+01,Walter+07}.
The \HI\ maps presented by these authors reveal one single \HI\ hole surrounding by a  supergiant shell. This structure, which comprises $\sim$75\% of the total \HI\ content of the galaxy, has a diameter of  $\sim$1.7~kpc and covers about half the optical extent of Ho~I. The supergiant shell found in Ho~I is explained by the strong stellar winds and supernova explosions within the starburst. It seems that part of the neutral gas has been lost to the intergalactic medium. But the agreement between the maps shown by \citet{Ott+01} in Ho~I and those obtained here for NGC~5253 is rather small. Furthermore,  the star-formation activity is much higher in NGC~5253 than in Ho~I. 
Both facts strongly suggest that NGC~5253 and Ho~I are experiencing different star formation histories.

Outflows seem to be accompanied by both \Ha\ and X-ray emission. 
For example, \citet*{SkillmanCMa03} reported filaments of \Ha\ emission in NGC~625 (another BCD and WR galaxy),
which are related to blowouts as they trace the path of hot gas into the halo \citep{Heiles93}. Effectively, the soft X-ray map of  NGC~625 \citep{BomansGrant98} reveals hot gas emission ($T\sim10^6$~K) which is associated with the ionized and neutral diffuse gas of the outflow. However, the `\HI\ SE plume' found in NGC~5253 is not associate with any \Ha\ or X-ray emission. Actually, the X-ray emission extends preferentially to the SW of the nucleus in this galaxy 
\citep[KS08]{Summers04}, casting doubt on its origin as an outflow.



NGC~625 is another BCDG hosting a large number of WR stars.
The ATCA \HI\ moment maps of NGC~625 presented by \citet{Cannon04} reveals that this galaxy also has perturbed neutral gas kinematics. 
These authors explain this following a blowout scenario which is the result of an spatially and temporally extended star formation event that the galaxy has experienced in the last 100~Myr. However, \citet{Cannon04} reported that NGC~625 is indeed undergoing solid-body rotation about its optical minor axis, as expected for an inclined rotating disc. 
Interestingly, their high- and low-resolution \HI\ maps of NGC~625 
agree well between them. However, the situation in NGC~5253 is different.  The high-resolution \HI\ maps are revealing the densest neutral gas in the galaxy, which is basically associated with the `NW \Ha+\HI\ shell' and the `SE \HI\ plume'. However, the low-resolution \HI\ map shows the diffuse neutral gas, which is mainly perturbed by external factors.


It is interesting to compare NGC~5253 with IC~10, which is the only known starburst within the Local Group.
As NGC~5253, IC~10 is a low-metallicity  --\abox=8.26, \cite{Skillman89, Garnett90, MagriniGoncalves09}-- low-mass galaxy showing 
a high star-formation activity \citep[i.e.,][]{HL90, Leroy06},
high far-infrared emission \citep{MelisseIsrael94} and non-thermal radio-continuum emission \citep{YangSkillman93}.
The outer \HI\ gas of IC~10 was discovered by \citet{ShostakSkillman89} and studied further by \citet{WilcotsMiller98}.
These authors characterized the neutral gas of IC~10 by a rotating disc 
embedded in an extended, clumpy and complex distribution of gas. 
\citet{WilcotsMiller98} also noted that the feedback of the massive stars has shaped shells and bubbles in the neutral gas of IC~10. 
However, the complex distribution and kinematics of the extended neutral gas 
around IC~10 suggest that this galaxy is still accreting gas from the large, extended \HI\ reservoir,
and hence the starburst may have been probably triggered by this gas falling in. 

Another interesting dwarf starbursting galaxy is NGC~1705. \citet*{Meurer98} 
presented an analysis of the \HI\ gas in this BCDG. They found an \HI\ rotating disc structure, but also a neutral gas distribution that appears disturbed. They reported the detection of an `\HI\ spur', which has around the 8\% of the total \HI\ mass of the galaxy, extending at least 4.5~kpc to the halo. These authors explained this feature as a consequence of an outflow, and then $\sim2\times10^7$~\Mo\ of neutral ISM has been expelled from the galaxy. This scenario agrees with the outflow of ionized gas powered by a central massive star clusters \citep{Marlowe95}.
\citet{Meurer98} also considered that the interaction origin of this feature, which may be an independent \HI\ cloud that is in process of merging with NGC~1705. They discarded this hypothesis because the optical appearance of the galaxy is not disturbed at all. As in NGC~1705, the optical distribution of NGC~5253 is also quite symmetric. 
However, in both cases the detection limit of the available optical images is $\mu_B\sim24-25$~mag\,arcsec$^{-1}$  \citep{LaubertsValentijn89,GildePaz03},
but the stellar features that may be associated with these \HI\ spurs may be much fainter \citep*[e.g.,][]{LSEGR06,LSE08}
and hence very deep images ($\mu_B\sim27-28$~mag\,arcsec$^{-1}$) are needed to detect these very faint features 
\citep[e.g.,][]{Martinez-Delgado09,Martinez-Delgado10}. Hence, we should not completely discard the hypothesis that the ` \HI\ spur' detected by \citet{Meurer98} in NGC~1705 actually has a tidal origin. A fundamental difference with NGC~5253 is that the  ` \HI\ spur'  found in NGC~1705  is not dominating the \HI\ dynamics of the galaxy, which follow a well-defined rotation pattern. But that is not the case of NGC~5253, where the effect of the `SE \HI\  plume' is completely masking the expected rotation of the galaxy and it is strongly perturbing the dynamics of the BCDG.



 \citet{vanEymeren10} reported evidence of merging in the BCDG IC~4662, as its diffuse \HI\ gas is somewhat distorted and it shows a tail of neutral gas. Indeed, spectroscopic observations \citep*{Hidalgo-GamezMO01,LS+11}
seem to reveal important physical and chemical differences in the ionized gas within  this galaxy, which also indicate that this galaxy actually is composed by two entities in interaction. Hence, in IC~4662 we found both outflows of neutral and ionized gas surrounding the star-forming regions but also distortions in the neutral gas that seem to be originated by a galaxy merger.

Finally, we compare NGC~5253 with the BCDG NGC~1569, which also hosts a strong starburst
\citep{Waller91,Gonzalez-DelgadoLHC97,BuckalewKobulnicky06}. This galaxy
seems that recently interacted with an independent \HI\ cloud \citep{StilIsrael98, StilIsrael02, Muhle05}.
It shows a rotation velocity of only 30~\kms \citep{Heckman95} but outflows with expansion velocities up to 100~\kms\ 
\citep*{Westmoquette+08}. The situation in this galaxy is quite similar to that found in NGC~5253: a perturbed \HI\ morphology and kinematics, high dispersion velocities of the central neutral gas, an almost inexistent rotation pattern, the probable detection of tidally ejected material, and outflows of both neutral 
\citep{StilIsrael02} and ionized \citep{Westmoquette+07} gas related to the violent star-formation activity in the starburst.
Hence we conclude that the main peculiar characteristic of the neutral gas in NGC~5253 (i.e., the apparent rotation about its optical major axis following the `SE \HI \ plume') is naturally explained by the infall of an independent \HI\ cloud and not by an outflow.

Summarizing, the comparison between NGC 5253 and a galaxy clearly hosting an outflow (Ho~I) indicates that the nature of the \HI\ gas has different origins.
 Furthermore, there is no \Ha\ or X-ray emission associated with the `SE \HI\ plume'.
However, the situation of NGC~5253 is similar to that observed in galaxies which seem to be in interaction with diffuse \HI\ clouds (IC~10, IC~4662, NGC~625, NGC~1705) and it is almost identical to that seen in a galaxy which is certainly known to be experiencing an interaction event with an independent \HI\ cloud (NGC~1569).


\subsection{Environment of NGC~5253}

NGC~5253 is located in
the Centaurus~A Group, which is one of the nearest galaxy groups within the Local Volume.
It consists of two subgroupings \citep{Karachentsev02,Karachentsev07}, one around Cen A (NGC~5128, which lies at 3.77~Mpc) and the other around M~83 (NGC~5236, which is at a distance of 4.6~Mpc). 
This late-type spiral galaxy possesses an unusually large \HI\ envelope \citep{Koribalski05,LVHIS} 
and it is surrounded by some dwarf irregular galaxies: ESO~444-G084, IC~4316, UGCA~365 (ESO~444-G078), NGC~5264 and NGC~5253. 
NGC~5253 lies at an angular distance of $\sim 1^{\circ}$\,54' and $\sim 12^{\circ}$ from M~83 and Cen~A, respectively. 

However, its distance obtained using different methods give different results
\citep[i.e.,][]{Saha95, Gibson00,Karachentsev02, Karachentsev07, Thim03, Sakai04,Davidge07, MouldSakai08}
ranging between 3.0 and 4.2~Mpc. Here we adopted 4.0~Mpc following \citet{KKHM04}. Hence, NGC~5253 is somewhat located at the periphery of the Cen~A\,/\,M~83 complex and between both subgroups. \citet{Karachentsev07}, however, located NGC~5253 in the Cen~A subgroup, but this galaxy is not far away from ESO~444-G084, IC~4316 and NGC~5264, which lie at around 4.5~Mpc, all belonging to the M~83 subgroup. The estimations of the distance to M~83 range between 4.5 and 4.9~Mpc, here we consider it is at 4.6~Mpc \citep{Karachentsev02}.


The morphology and kinematics of the \HI\ gas within NGC~5253 are much more difficult to explain than that seen in the nearby galaxies IC~4316, ESO~444-G084 and NGC~5264 (see Koribalski et al. 2011 to see their \HI\ maps). These three galaxies mainly show symmetric \HI\ morphologies and a pattern of rotation, although some mild disturbances seem to be found in ESO~444-G084 and NGC~5264.
In the latter case 
there is some evidence that the galaxy interacted with an independent object in the past.
Current cosmological models indicate that purely isolated galaxies should be extremely rare. Indeed,
we should expect to find 
distortions in the outer parts of the galaxies and, in some way, it should be very difficult to find a `pure" isolated system \citep{Koribalski10}. These disturbances reveal the processes that are building up the galaxies and are very evident in strong star-forming dwarf systems when deep data are available 
\citep[e.g.,][]{LSE08,LSE09,LS10}.  
The peculiar morphology and kinematics of the \HI\ gas reported here in NGC~5253 and the very intense starburst that this galaxy is experiencing strongly suggest a connection between both galaxy properties.

It is interesting to note that NGC~5253 is also close to the late-type spiral galaxy M~83, which lies at a radial distance of only $\sim$600~kpc,  being the projected separation between them $\sim$130~kpc. LVHIS data of this galaxy  \citep{Koribalski05,LVHIS}  are remarkable, as its \HI\ distribution (which has a size of $\sim$100~kpc, i.e., at least 5 times larger than its optical Holmberg diameter) has streamers, irregular enhancements, and more interesting, an asymmetric tidal arm which points out to the direction of NGC~5253. This tidal structure indicates that M~83 has accreted or strongly interacted with a dwarf galaxy in the past. 
Furthermore, \citet{Bresolin+09} reported an abrupt discontinuity in the radial oxygen abundance trend near the optical edge of the disc of M~83, which they also associated to a past galaxy encounter.
Could it be with NGC~5253? Previous authors \citep{vdBergh80,CaldellPhillips89} already 
suggested that the nuclear starburst observed in NGC~5253 was triggered as a result with an interaction with M~83 around 1~Gyr ago. 
In this sense, the main disturbances in the \HI\ distribution we observe in NGC~5253 are precisely located in the direction to the long tidal arm found in M~83.
Hence, our data may suggest that a diffuse \HI\ filament between both galaxies exist, and part of it may be infalling NGC~5253. Indeed, the metallicity computed by \citet{Bresolin+09} in the external disc of M~83 --\abox$\sim$8.2-- is a good match to that estimated in NGC~5253. 
This diffuse gas stripped from the external \HI\ disc of M~83 may be the origin of the 
\HI\ cloud that is infalling along the minor axis of NGC~5253.
The fact that the real duration of the starburst in NGC~5253 may be longer than 450~Myr \citep{McQuinn+10a,McQuinn+10b} ties very well with either the ongoing infall that the galaxy may be experiencing now or the effects with an interaction with M~83 that started well before the currently observed star-formation burst.

\subsection{Other peculiarities of NGC~5253}

\begin{figure}
\includegraphics[angle=-90,width=\linewidth]{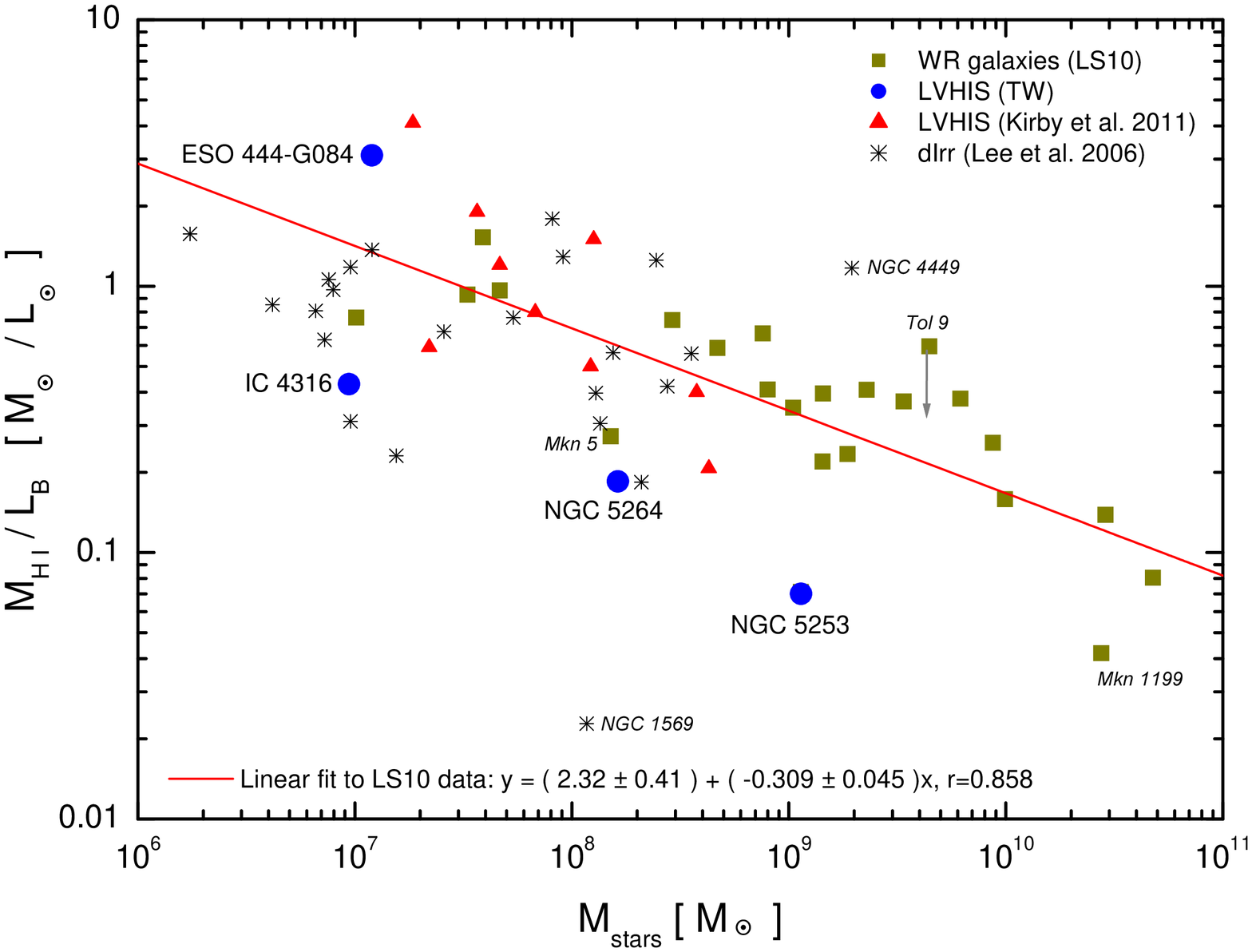} 
\caption{\label{mhil_mstars} Comparison between the stellar mass, \Mstars, and the \HI\ mass-to-light ratio for NGC~5253 and its nearby dwarf galaxies (blue circles). We compare with the sample of nearby dwarf irregular galaxies analyzed by \citet{LeeZG07}, with the sub-sample of the LVHIS galaxies studied by Kirby et al. (2010) (red triangles), and with the sample of strong star-forming (Wolf-Rayet) galaxies  analyzed by \citet{LS10} (dark yellow squares), which included NGC~5253. The red solid line shows a linear fit to the \citet{LS10}  data.}
\end{figure}

\citet{LS10} analyzed a sample of strong star-forming galaxies (Wolf-Rayet galaxies) that included NGC~5253. He found that, contrary to that observed in starburst galaxies and in star-forming regions or $UV$-rich star clusters in spiral galaxies  \citep{K07,KoribalskiLS09}, NGC~5253 does not obey the Schmidt-Kennicutt (S-K) scaling laws of star-formation \citep{Schmidt59, Schmidt63,K98}.

\citet{LS10} also compared the \HI\ mass-to-light ratio (which is a distance-independent quantity) with the total stellar mass, \Mstars, 
for his galaxy sample (see his Fig~16). 
We reproduce this plot in our Fig.~\ref{mhil_mstars}, his data points are the dark yellow squares but adding the position of the LVHIS galaxies analyzed by \citet{Kirby10} (red triangles) and
the data points of the sample of nearby dwarf irregular galaxies studied by \citet{LeeZG07} 
(stars). 
As noticed by \citet{LS10}, the position of NGC~5253 in this diagram clearly disagrees with the position of the other galaxies, as it shows a \MHil\ ratio that is an order of magnitude smaller than that expected from its stellar mass. 
We note that NGC~1569, that shares similar properties to those observed in NGC~5253 (including a probable interaction with an independent \HI\ cloud, see Sect~5.2), also lies very far from the position of the other galaxies  in the diagram plotted in Fig.~\ref{mhil_mstars}.

\citet{LS10} also compared the \Mg/\Mstars\ ratio and the oxygen abundance for his galaxy sample. His Fig.~17 clearly shows that the \Mg/\Mstars\ ratio decreases with increasing metallicity, indicating that the importance of the stellar component to the total mass is larger in more massive galaxies.
Indeed, galaxies show a clear correlation between the oxygen abundance and the total stellar mass (see left panel of Fig.~14 in L\'opez-S\'anchez 2010 and Lara-L\'opez et al. 2010). 
Again, the position of NGC~5253 in this diagram does not agree with that observed for the rest of the galaxies, as it seems to have a $\sim$5 times lower \MHil\ ratio than that expected for its metallicity.
This may indicate that the oxygen abundance derived in the central, 
intense star-forming region is not reflecting the metallicity of the stellar component of the galaxy (i.e., the metallicity of stars born in previous star-formation events), which may be higher than that derived from the analysis of the ionized gas. Assuming the derived \Mg/\Mstars\ ratio for NGC~5253, we should expect that this galaxy has a metallicity up to 0.3--0.4 dex higher than that computed in the central starburst [\abox=8.18$\pm$0.03, L\'opez-S\'anchez et al. 2007]. Hence, the infall of low-metallicity neutral gas that we suggest here may have lowered the gas-phase abundance of this galaxy. Interestingly, \citet{LSEGRPR07}  found that the youngest starbursts --regions A and B, with \abox=8.18$\pm$0.03-- have a metallicity which is 
$\sim$0.1 dex lower than that reported in slightly older areas --regions C and D, with \abox=8.30$\pm$0.05--. 
%
%
%
%

\citet{KoppenHensler05} created chemical evolution models that consider massive and rapid accretion of metal-poor gas into a galaxy. They found that the oxygen abundance is decreased during the infall due to dilution of the galactic gas, and that gas masses of 10$^6$~\Mo\ 
would be sufficient to cause a large influence on dwarf galaxies.
They concluded that such a collision could leave observable marks in the chemical properties of the galaxies, different from what a closed-box evolution predicts.
Indeed, NGC~5253 do not follow a close-box model \citep[see Fig.~21 in][]{LS10}, as it happens in the majority of the WR galaxies analyzed by this author. Actually, all the predictions indicated by \citet{KoppenHensler05}  are satisfied in the case of NGC~5253, who also 
predicted that the galaxy may be very bright if the infall triggers enhanced star-formation. 
Hence, the hypothesis of the infall of an \HI\ cloud in NGC~5253 will naturally explain its relatively high baryonic mass for its oxygen abundance \citep[see left panel of Fig.~14 in][]{LS10}, its very low \MHil\ ratio 
 (and its position in Fig.~\ref{mhil_mstars}) 
and its deviation of the S-K scaling laws of star-formation. 


\section{Conclusions}


We have presented deep \HI\ line and 20-cm radio continuum data of the intriguing blue compact dwarf galaxy NGC~5253.
The data were obtained using the ATCA as part of the LVHIS survey, but also from our multiwavelength analysis of BCDGs.
Hence, we completed our study using multiwavelength data extracted from the literature, 
which includes X-ray, $FUV$, optical $B$- and $R$-band, \Ha, \mbox{$NIR$ $H$-band,} and $FIR$ data.

Our deep low-resolution \HI\ maps show, for the first time, the very disturbed \HI\ morphology that NGC~5253 possesses in its external areas, which includes tails, plumes and detached \HI\ clouds.  Our data recovers almost 96\% of the single-dish \HI-flux, indicating that very few diffuse emission has been filtered out in our interferometric observations. We derive a total \HI\ mass of $1.59\times10^8$\,\Mo\ in NGC~5253, around 5\% of this mass ($\sim8\times10^6$~\Mo) is located in the diffuse almost detached gas found at the northern area of the galaxy. Adding the stellar mass to the neutral gas mass, we estimate that the baryonic mass of NGC~5253 is $13.5\times10^8$\,\Mo.

The high-resolution \HI\ maps trace an \Ha\ shell at the NW of the galaxy, which has a mass of $\sim5.4\times10^6$~\Mo, and it is related to an expanding bubble in the ISM that
seems to be a consequence of the combined actions of the winds of massive stars and supernova explosions. 
However, this expanding bubble will almost certainly not end in a galactic wind.
Our high-resolution \HI\ maps also confirm the discovery of a `SE \HI\ plume' made by \citet{KS08}, 
 which is now clearly detected. This feature is not associated with any \Ha\ or X-ray emission and hence it is most likely not related to an outflow.
Indeed, the kinematics of the neutral gas in NGC~5253 derived from the low- and medium-resolution \HI\ maps 
is highly perturbed and do not follow a rotation pattern. 
We argued that this feature actually is an independent \HI\ cloud which is interacting with the main body of the galaxy.
However, the \HI\ data may also suggest that some of the features observed in NGC~5253 have been originated as the result of ram pressure stripping from moving through a dense IGM.


Our 20-cm radio-continuum maps show that the emission of NGC~5253 is resolved and 
associated with the intense star-forming region located at the center of the galaxy. 
We estimate the star-formation rate ($SFR$) 
using a multiwavelength approach, 
and compare the results with other galaxy properties. 
NGC~5253 does not satisfy the Schmidt-Kennicutt law of star-formation, has a very low \HI\ mass-to-light ratio when comparing with its stellar mass, and seems to be slightly metal-deficient in comparison with starburst galaxies of similar baryonic mass.
The relatively high stellar mass also indicates that the BCDG has experienced a relatively long star-formation history.

We have compared the properties observed in NGC~5253 with those seen in three nearby galaxies members 
of the M~83 subgroup (IC~4316, ESO~444-G084 and NGC~5264, for which we have also derived their properties 
using the data provided by the LVHIS survey) and with the properties found in other starbursting dwarf galaxies. 
%

Taking into account all available observational data and the comparison with similar starbursting galaxies, we conclude that NGC~5253 is very probably experiencing the infall of a diffuse, low-metallicity \HI\ cloud along the minor axis of the galaxy, which is comprising the ISM and triggering the powerful central starburst.
Both entities seem to be now in a process of merging, but the interaction probably started some time ago, as we detect tidal gas at the northern regions of the galaxy and it seems that previous starbursts 
have occurred in the system. 
Perhaps the origin of this low-metallicity \HI\ cloud is related to the very asymmetric tidal \HI\ arm detected in M~83, which may indicate that both galaxies have experience a strong interaction in the past.

This study reinforces the conclusions reached by \citet{LS06,LS10},  
which suggests that interactions with or between dwarf galaxies or \HI\ clouds trigger the star-formation activity in many dwarf and normal starburst galaxies.

Following the proposed infall scenario, perhaps the combined actions of the accretion of
low-metallicity gas and the outflow found at the NW of the galaxy 
(which is now moving the metal-rich material into the surrounding IGM)
have lowered the total oxygen abundance of NGC~5253,  
as its mass-to-light and mass-to-mass ratios seem to match better with an object with metallicities up to 0.3--0.4 dex higher than those computed in the youngest starburst.
Hence, we conclude that detailed and comprehensive multiwavelength analysis, such as this presented
here, is a powerful tool to elucidate the complex relationships between star-formation activity and
the structure of the inner and external ISM, as well as to constrain the evolution and fate of the gas
in star-forming galaxies.

\section*{Acknowledgements}
We thank the referee, Evan D. Skillman, for his detailed reviewing, that improved the quality  of this paper.
Based on observations made with the ATCA
(Australia Telescope Compact Array), which is funded by the Commonwealth of Australia for operation as a National Facility managed by CSIRO. 
This research has made extensive use of the NASA/IPAC Extragalactic 
      Database (NED) which is operated by the Jet Propulsion Laboratory, 
      Caltech, under contract with the National Aeronautics and Space 
      Administration. 
     The Digitised Sky Survey was produced by the Space Telescope Science
      Institute (STScI) and is based on photographic data from the UK Schmidt 
      Telescope, the Royal Observatory Edinburgh, the UK Science and 
      Engineering Research Council, and the Anglo-Australian Observatory.      
The  Galaxy  Evolution  Explorer (GALEX) 
is  a  NASA Small  Explorer,  launched  in  April 2003.  We  gratefully  acknowledge NASA's  support  for  construction, 
operation, and science analysis for the GALEX mission. 
 This research has made extensive use of the
SAO/NASA Astrophysics Data System Bibliographic Services (ADS).






\end{document}